\documentclass[journal]{IEEEtran}
\newcommand{\RNum}[1]{\uppercase\expandafter{\romannumeral #1\relax}}
\usepackage{mathrsfs}
\usepackage{amssymb}
\usepackage[mathcal]{euscript}
\usepackage{diagbox}
\usepackage{cite}
\usepackage[T1]{fontenc}
\usepackage{graphicx}
\usepackage{CJKutf8}
\usepackage{xspace}
\usepackage{makecell}
\usepackage{psfrag}
\usepackage{url}
\usepackage{stfloats}
\usepackage{amsmath}
\usepackage{array}
\usepackage{float}
\usepackage{multirow} 
\usepackage{hyperref}
\usepackage{amsthm}
\usepackage{color}
\usepackage{multicol}
\usepackage{stfloats}
\usepackage{enumerate}
\usepackage[utf8]{inputenc} 
\usepackage[T1]{fontenc}    
\usepackage{hyperref}       
\usepackage{url}            
\usepackage{booktabs}       
\usepackage{amsfonts}       
\usepackage{nicefrac}       
\usepackage{microtype}      
\usepackage{color}
\usepackage{xcolor}
\usepackage{graphicx}
\usepackage{multirow}
\usepackage{flushend}
\usepackage{soul}
\usepackage{mathrsfs}
\usepackage{bbm}
\usepackage{amssymb}

\usepackage[mathcal]{euscript}

\usepackage{psfrag,calc,url,bm}

\usepackage{cite}

\usepackage{graphicx}

\usepackage{psfrag}

\usepackage{subfigure}

\usepackage{url}

\usepackage{stfloats}

\usepackage{amsmath}

\usepackage{float}

\usepackage{algorithmic}

\usepackage[ruled,linesnumbered,vlined]{algorithm2e}

\usepackage{color}

\usepackage{boxedminipage}

\usepackage{amsthm}

\usepackage{multirow}

\usepackage{setspace}

\usepackage{array}
\usepackage{wrapfig}
\usepackage{bbm}
\usepackage{mathtools}
\usepackage{dsfont}
\colorlet{soulgray}{lightgray!30}
\sethlcolor{soulgray}
\usepackage{subfigure}
\usepackage{booktabs}  
\newtheorem{remark}{Remark}

\newtheorem{definition}{Definition}
\newtheorem{proposition}{Proposition}
\allowdisplaybreaks[4]
\usepackage[ruled]{algorithm2e}
\usepackage{algorithmic} 
\usepackage{fancyhdr}
\ifCLASSINFOpdf
\else
\fi
\newcommand{\doc}{\ensuremath{z}}

\newcommand{\answer}{\ensuremath{m}}
\newcommand{\genparam}{\ensuremath{\theta}}
\newcommand{\retparam}{\ensuremath{\eta}}
\newcommand{\query}{\ensuremath{q}}

\newcommand{\denc}{\mathbf{E}_{\rm z }}
\newcommand{\qenc}{\mathbf{E}_{\rm q}}

\begin{document}
\begin{CJK}{UTF8}{gbsn}
%
\title{Generative AI Agents with Large Language Model for Satellite Networks via a Mixture of Experts Transmission}


%
	\author{Ruichen Zhang, Hongyang Du, Yinqiu Liu, Dusit Niyato,~\IEEEmembership{Fellow,~IEEE},  Jiawen Kang, \\ Zehui Xiong,   Abbas Jamalipour,~\IEEEmembership{Fellow,~IEEE}, and Dong In Kim,~\IEEEmembership{Fellow,~IEEE}




\thanks{R. Zhang, H. Du, Y. Liu, and D. Niyato are with the College of Computing and Data Science, Nanyang Technological University, Singapore (e-mail: ruichen.zhang@ntu.edu.sg, hongyang001@e.ntu.edu.sg, yinqiu001@e.ntu.edu.sg, dniyato@ntu.edu.sg).}

\thanks{J. Kang is with the School of Automation, Guangdong University of Technology, China (e-mail: kavinkang@gdut.edu.cn).}

\thanks{Z. Xiong is with the Pillar of Information Systems Technology and Design, Singapore University of Technology and Design, Singapore (e-mail: zehui\_xiong@sutd.edu.sg).}

\thanks{A. Jamalipour is with The University of Sydney, Sydney NSW 2006, Australia (e-mail: a.jamalipour@ieee.org).}

\thanks{D. I. Kim is with the Department of Electrical and Computer Engineering, Sungkyunkwan University, Suwon 16419, South Korea (email:dongin@skku.edu).}

}
\maketitle

\begin{abstract}
In response to the needs of 6G global communications, satellite communication networks have emerged as a key solution. 
However, the large-scale development of satellite communication networks is constrained by the complex system models, whose modeling is challenging for massive users.
Moreover, transmission interference between satellites and users seriously affects communication performance. 
To solve these problems, this paper develops generative artificial intelligence (AI) agents for model formulation and then applies a mixture of experts (MoE) approach to design transmission strategies. 
Specifically, we leverage large language models (LLMs) to build an interactive modeling paradigm and utilize retrieval-augmented generation (RAG) to extract satellite expert knowledge that supports mathematical modeling.
Afterward, by integrating the expertise of multiple specialized components, we propose an MoE-proximal policy optimization (PPO) approach to solve the formulated problem. Each expert can optimize the optimization variables at which it excels through specialized training through its own network and then aggregates them through the gating network to perform joint optimization. The simulation results validate the accuracy and effectiveness of employing a generative agent for problem formulation. Furthermore, the superiority of the proposed MoE-ppo approach over other benchmarks is confirmed in solving the formulated problem. The adaptability of MoE-PPO to various customized modeling problems has also been demonstrated.
\end{abstract}


\section{Introduction}
In the era of 6G, the demand for global communication continues to increase rapidly, spotlighting the significant role of satellite communications, particularly due to their unparalleled advantages in overcoming the constraints faced by ground communication systems \cite{8368236}. Among these, low-earth orbit (LEO) satellite communication networks have emerged as a pivotal solution to the limitations of ground communication systems, offering high-speed, low-latency communication services with extensive coverage and enhanced security \cite{9502642}.  This makes LEO satellite systems critical for ensuring connectivity in remote areas such as oceans and mountainous areas that are difficult to reach with traditional communications systems \cite{9749193}. Moreover, with an increasing number of satellite users, there is an urgent need for increased capacity in satellite communication networks \cite{9210567}. This surge in demand underscores the importance of advancing and refining these systems to meet future communication requirements, presenting researchers with the challenge of scaling and optimizing satellite communications to address these evolving requirements \cite{9403416}. However, to meet these demands, satellite communication networks face two main challenges as follows.
\begin{itemize}
    \item \textbf{Challenge  \uppercase\expandafter{\romannumeral1}:} The mathematical modeling of satellite communications is much more complex than that of ground communications systems \cite{10209551}. It is necessary to consider the curvature of the earth, the impact of the atmosphere on the signal, and non-uniformity of communications traffic \cite{345892}. This complexity makes satellite scenarios (i.e., homogeneous or heterogeneous systems), channel modeling (i.e., static or dynamic channels), access protocols (i.e., rate-splitting multiple access or space-division multiple access), optimization goals (i.e., energy efficiency (EE) and spectral efficiency (SE)) becomes extremely challenging, which not only sets a high threshold for newcomers but also poses a challenge to interdisciplinary researchers who need to understand the operating mechanism of satellite communications.
    
   \item \textbf{Challenge  \uppercase\expandafter{\romannumeral2}:}{Compared with ground communication systems, satellite communication networks face more complex challenges in resource allocation \cite{10413484}. Due to the wide coverage of satellite communications and the large space where users are distributed, coupled with the limited resources and fixed beam coverage design of satellites, it is difficult to meet the diverse and varying requirements of different kinds of users \cite{9371230}.} Moreover, transmission interference between satellites, between beams, and between users seriously affects communication performance \cite{9970355}. Therefore, developing adaptive resource allocation schemes to mitigate interference and manage resources effectively is critical to improving the quality of service (QoS) of satellite communication networks.
\end{itemize}

To address Challenge \uppercase\expandafter{\romannumeral1}, an interactive generative artificial intelligent (AI) agent has been introduced \cite{park2023generative}. Generative AI Agents can integrate the power of large language models (LLMs) such as ChatGPT and LLaMA with retrieval-augmented generation (RAG) technologies, aiming to generate solutions to specific problems through interactive sessions with human users \cite{zhang2024interactive}. Specifically, the LLM is able to respond by understanding the deep meaning and context of natural language and generating detailed answers to specific questions \cite{du2023user}. RAG is able to retrieve relevant information from massive documents, providing LLMs with real-time, rich background knowledge, allowing LLMs to provide more accurate and informative content when answering queries \cite{lewis2020retrieval}. In the context of optimizing satellite communication networks, 
the interactive approach enables the generative AI agent to deeply understand and adapt to specific application scenarios and requirements \cite{zhang2023generative1}. For example, when faced with modeling challenges, including changes in satellite orbit parameters, signal propagation characteristics, or complex interference conditions, generative AI agents can refine a problem definition and collect necessary data to generate a more accurate and adaptive system model by interacting with human users and network environments. 

To address Challenge \uppercase\expandafter{\romannumeral2}, the mixture of experts (MoE) model has been introduced. MoE integrates multiple specialized deep neural network (DNN) components, i.e., experts, to handle complex tasks \cite{6215056}. {Specifically, each expert is fine-tuned and specialized for a specific type of sub-task or sub-dataset, providing expertise in efficient problem-solving capabilities within their domain, which can be more effective than monolithic models \cite{masoudnia2014mixture}.} MoE uses a gating network technique to divide larger tasks into smaller, more manageable units and dynamically select the most appropriate combination of experts to handle these tasks based on the input data characteristics. Unlike traditional neural network models, MoE selectively activates a subset of experts based on specific features of the input, thereby improving the model's processing efficiency and performance \cite{chen2023mod}. {In satellite communication networks, MoE can provide customized solutions for various challenges based on different experts, such as spectrum allocation, signal interference reduction, and energy efficiency optimization. For example, under dynamically changing communication environments and diverse requirements of different users, MoE can achieve effective resource allocation and interference management by selecting a combination of experts that are most suitable for handling current signal conditions and user distribution \cite{ACM}.}

Inspired by the above discussion, as shown in Fig.~\ref{Architecture_v1}, this paper first studies a framework for modeling using generative AI agents and then optimizes the transmission strategy of satellite communication networks through an MoE model. {For the generative AI agents, unlike traditional system modeling, which typically relies on the designer's knowledge and experience to manually formulate mathematical models, our generative AI agents leverage the vast knowledge embedded in LLMs and the precise contextual information provided by RAG to customize and adapt the modeling process to specific satellite communication needs. For the transmission strategy, instead of using a single comprehensive solution system to optimize all variables or employing a multi-agent system to address different aspects, our proposed MoE approach utilizes multiple specialized experts, each focusing on optimizing specific variables, thereby improving efficiency.}
\textit{To the best of the authors' knowledge, this is the first work in the networking field to adopt generative AI agents for customized modeling and then solve the formulated problem by MoE.\footnote{https://github.com/RickyZang/GAI-agent-satellite}}  The contributions of this work are summarized as follows.
\begin{itemize}
   \item We introduce a generative AI agent framework to customize system problem formulation. This approach effectively addresses the diverse modeling requirements within satellite communication networks via a two-layer semantic router and RAG. The framework leverages the rich knowledge of LLMs and the satellite expertise derived by RAG to provide precise modeling solutions for specific problems in satellite communication networks. (For Challenge \uppercase\expandafter{\romannumeral1})
    \item {Based on different satellite scenarios, we employ different access strategies, channel models, and optimization objectives for heterogeneous and homogeneous satellite networks through the generative AI agent framework to customize our system modeling.} This step ensures that the models accurately reflect the complexity and specific requirements of the actual communication environment in both LEO and Geostationary Earth Orbit (GEO) satellite networks. {By doing so, we enable a more efficient selection of modeling parameters based on designers' requirements, reducing the risk of errors and saving time.} (For Challenge \uppercase\expandafter{\romannumeral1}) 
 \item  We propose an MoE-based proximal policy optimization (PPO) method to optimize the formulated problem generated by the generative AI agent. {Specifically, each expert can optimize the optimization variables for which it has been trained through its network and then aggregate them through the gating network to perform joint optimization, which can maximize the efficiency of spectrum resource utilization while meeting communication quality requirements.} (For Challenge \uppercase\expandafter{\romannumeral2})
\end{itemize}

The rest of this paper is organized as follows. The related work is reviewed in Section \ref{work}. Section \ref{Framework} shows our proposed Generative AI Agent framework. An MoE-based PPO approach is proposed in Section \ref{PPO}. Section \ref{sim} conducts some simulations of the proposed generative AI agent framework and MoE-based PPO approach. Finally, we conclude the whole work in Section \ref{con}.

\section{Related Work}\label{work}
In this section, we review literature across three domains, i.e., satellite communication networks, Generative AI Agents, and the MoE paradigm, highlighting the advancements and identifying gaps that our work aims to bridge.

\subsection{Satellite Communication Networks}
Recent advancements in satellite communication networks have garnered significant attention due to their potential benefits of global connectivity. {For instance, Chen et. al. \cite{9749193} proposed federated learning in LEO satellite communication networks to support massively interconnected devices and reduce communication overheads, showing through simulations that the method significantly reduces latency and improves satellite communication performance in 6G systems.} Also, in \cite{9419053}, an optimization problem for maximizing the sum rate in massive MIMO LEO satellite networks, considering both imperfect successive interference cancellation (SIC) and channel state information (CSI). Similarly, Huang et. al. \cite{huang2023deep} addressed the sum rate maximization problem in LEO satellite-terrestrial networks utilizing RSMA and proposed a PPO-based approach to meet diverse QoS requirements. Khan et. al. \cite{khan2023rate} focused on sum rate maximization in heterogeneous satellite networks, introducing efficient subcarrier beam assignment and power allocation strategies for secondary LEO satellites while managing interference with primary GEO satellites. Besides spectral efficiency, in \cite{8957364}, the system EE was maximized in integrated satellite-terrestrial radio access networks by jointly optimizing the cache sharing vector, block placement vector, and power allocation vector, where the information EH QoS requirements of users and the power budget are considered as constraints. Despite these contributions, the complex models inherent in these efforts may create barriers to understanding for newcomers to the field, which requires a framework that makes entry into this field of research more accessible.

\subsection{Generative AI Agent}
The field of generative AI has seen significant interest, particularly in its application to interactive agents \cite{du2024enhancing}.  For instance,  Wang et. al. \cite{wang2023does} developed an interactive chatbot leveraging LLMs to mimic character-specific personality traits, achieving a high correlation (i.e., about 82.8\%) with human perception. Also, The SWIFTSAGE framework, introduced by Lin et. al. \cite{lin2023swiftsage}, combines behavioral cloning with LLM prompting for enhanced action planning in complex tasks.  Furthermore,  generating agents are also used in the networking field. For instance,  Du et. al. \cite{du2023user} demonstrated a generative AI agent empowered by LLMs to simulate user interactions, offering insights into real-time Quality of Experience (QoE) feedback, where experiments showed that using the proposed agent can improve performance by 15\%. Also, Zhang et. al. \cite{zhang2024interactive} proposed an optimization framework including environment, action, brain, and perception by using an Generative AI Agent, where pluggable LLM and RAG modules are employed for decision-making knowledge base and contextual memory. However, these aforementioned works focus mainly on direct policy generation and do not consider the potential for customized and selective system modeling and network configuration.

\begin{figure}[!t]
\centering
\includegraphics[width=0.49\textwidth]{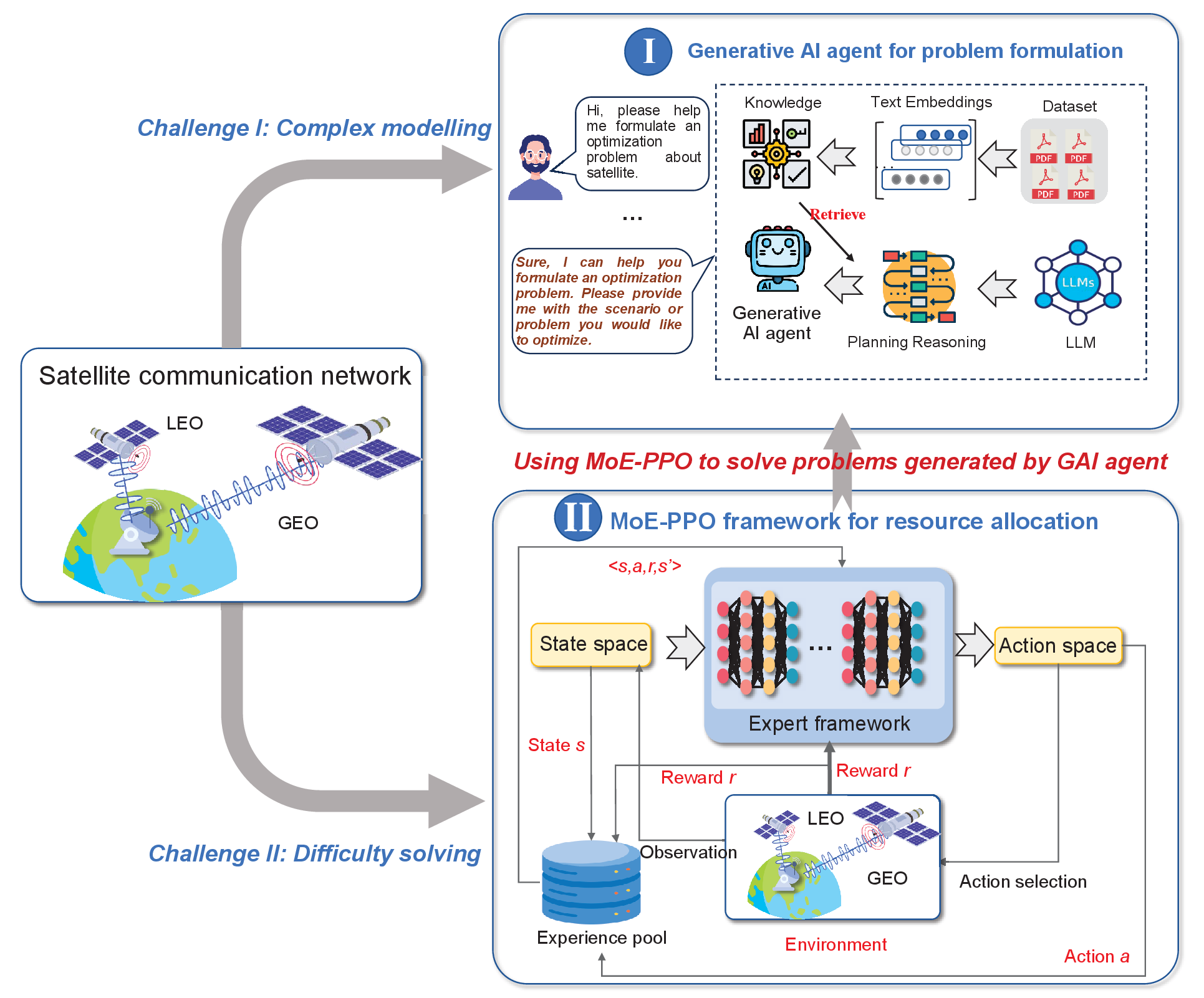}
\caption{The outline of the paper. Specifically, to address the challenges of complex modeling and difficulty in solving formulated problems within satellite communication networks, we introduce a generative AI agent and an MoE approach. The generative AI agent is tasked with formulating the optimization problems, while the MoE framework is employed to effectively solve the optimization problems formulated by the generative AI agent.}
\label{Architecture_v1}
\end{figure}

\subsection{Mixture-of-Experts}

The MoE model, initially introduced by Jacobs et. al. \cite{jacobs1991}, represented a shift in supervised learning, facilitating parallel processing by multiple specialized networks.
With the emergence of deep neural networks (DNN), Eigen et. al. \cite{eigen2013learning} proposed an architecture that can extend multiple experts in parallel on each neural network layer. Each expert has its own weight matrix, and then the gating network distributes the weights to multiple experts, requiring all experts to perform calculations during reasoning. In the era of large models, Du et. al. \cite{du2022glam} developed a large model combining LLM and MoE, where MoE can dynamically select only the most relevant experts for the current task to be calculated, thereby reducing the computation of floating-point operations. Thanks to these developments and advantages, MoE is also used in networking fields. For instance, Jaiswal et. al. \cite{jaiswal2023leveraging} leveraged MoE for a data-driven transfer learning approach, enhancing radio map models through a combination of location-based and independent expertise. Also,  Lopez et. al. \cite{lopez2020channel} applied an MoE-based learning scheme to joint source-channel coding (JSCC), demonstrating the MoE models' versatility and robustness. Nonetheless, existing works focus mainly on MoE to enhance model architecture rather than exploiting its potential for resource optimization in networking. Therefore, this is the primary focus of this work to fill the gap.

\section{Generative AI Agent Framework}\label{Framework}

As shown in Fig.~\ref{Architecture_v1}, to address Challenge \uppercase\expandafter{\romannumeral1}, this section proposes a generative AI agent to achieve customized model formulation.

\subsection{Dataset Construction}

Due to the curvature of the earth, the influence of the atmosphere on signals of satellites, mathematical modeling of satellite communications is difficult and complex, where main four aspects (i.e., scenarios, access protocols, channel models, and optimization goals) should be considered. Thus, in the process of customizing a satellite communication model, the first step is to establish a specialized database due to the lack of ready-made databases. To support subsequent problem formulation, we need to build such a database based on the above four aspects so that the generated AI agent can perform generation.



\subsubsection{Scenarios}
There are significant differences in satellite communication models in different scenarios. Therefore, we consider two typical scenarios, i.e., the homogeneous satellite communication model and the heterogeneous satellite communication model\cite{deng2019next}.

For the homogeneous scenarios, we consider a downlink LEO satellite and terrestrial network,  where an  $N_{\rm T}$-antenna LEO satellite serves $K$  single-antenna LEO ground users (LGUs).  Let ${\mathbb E}\{|s_k|^2\}=1$ where $k\in{\mathcal K}=  \{1,...,K\}$. Thus, the transmit signal is given by
\begin{flalign}\label{trans_signal}
{\bf x} = \sum\limits_{k=1}^{K}{\bf w}_k s_k,
\end{flalign}
where ${\bf w}_k \in \mathbb{C}^{N_{\rm T}\times 1} $ denotes the beamforming vector associated with the stream $s_k$.

For the heterogeneous scenarios, we consider a downlink heterogeneous satellite network composed of one $N_{\rm M}$-antenna GEO satellite and one $N_{\rm T}$-antenna LEO satellite, where these satellites use the same spectrum. The GEO satellite serves $M$ single-antenna GEO ground users (GGUs) and each LEO satellite serves $K$ LGUs. For clarity, Let ${\mathbb E}\{|s_k|^2\}=1$ and  ${\mathbb E}\{|s_m|^2\}=1$. We use $0$ to denote the index of the GEO satellite, $m \in{\mathcal{M}}=\{1,2,...,M\}$ to denote the index of the $m$-th GGU and $k \in{\mathcal{K}}=\{1,2,...,K\}$ to denote the index of the $k$-th LGU.  Thus, the transmit signals of GEO and LEO are, respectively, given by
\begin{equation}
\begin{aligned}
 \textbf{x}_{\rm G} =  \sum\limits_{m=1}^{M}\textbf{w}_{m}s_{m},
\end{aligned}
\end{equation}
and
\begin{equation}
\begin{aligned}
\textbf{x}_{\rm L} =  \sum\limits_{k=1}^{K}\textbf{w}_{k}s_{k}.
\end{aligned}
\end{equation}

\subsubsection{Access Protocols} In different satellite communication scenarios, the selected access protocols, such as SDMA and RSMA, will also have different impacts on system modeling. For example, when we choose the heterogeneous scenario, for SDMA, it allows to use spatial separation technology to allow the system to serve multiple ground users simultaneously. Thus, the corresponding SINR at the $m$-th GGU and the $k$-th LGU are, respectively, given by
\begin{equation}
\begin{aligned}
\Gamma_{m}\left(\{\textbf{w}_{k}\} \right)= \frac{\lvert \textbf{h}_{{\rm G},m}^{H}\textbf{w}_{m}\rvert ^{2}}{\sum\limits_{ m' \neq m}^{M}\lvert \textbf{h}_{{\rm G},m}^{H}\textbf{w}_{m'}\rvert^{2} + \sum\limits_{k=1}^{K}\lvert \textbf{h}_{{\rm L},m}^{H}\textbf{w}_{k}\rvert^{2} +\sigma_{a}^2},
\end{aligned}
\end{equation}
and
\begin{equation}
\begin{aligned}
\Gamma_{k}\left(\{\textbf{w}_{k}\} \right)= \frac{ \lvert \textbf{h}_{{\rm L},k}^{H}\textbf{w}_{k}\rvert ^{2}}{	 \sum\limits_{k'\neq k}^{K}\lvert \textbf{h}_{{\rm L},k}^{H}\textbf{w}_{k'}\rvert^{2} 
+\sum\limits_{m=1}^{M}\lvert \textbf{h}_{{\rm G},k}^{H}\textbf{w}_{m}\rvert^{2}+ \sigma_{b}^2},
\end{aligned}
\end{equation}
where $\textbf{h}_{{\rm G},m}$ and $\textbf{h}_{{\rm L},k}$ denote the direct channel vector between GEO and the $m$-th GGU  as well as channel vector between LEO and the $k$-th LGU, respectively. $\sigma_{a}^2$ and $\sigma_{b}^2$ denote the corresponding Gaussian noise at $m$-th GGU and $k$-th LGU, respectively.

For RSMA, it decomposes signals into common and private parts to improve system performance. To mitigate the co-channel interference in heterogeneous scenarios, the $1$-layer RSMA scheme is adopted at the LEO satellite \cite{clerckx2023primer}. Thus, the corresponding SINR at the $m$-th GGU is given by
\begin{equation}
\begin{aligned}
&\Gamma_{m}\left(\{\textbf{w}^{\rm p}_{k}, \textbf{w}^{\rm c}\} \right)\\
&= \frac{\lvert \textbf{h}_{{\rm G},m}^{H}\textbf{w}_{m}\rvert ^{2}}{\sum\limits_{ m' \neq m}^{M}\lvert \textbf{h}_{{\rm G},m}^{H}\textbf{w}_{m'}\rvert^{2} + \sum\limits_{k=1}^{K}\lvert \textbf{h}_{{\rm L},m}^{H}\textbf{w}^{\rm p}_{k}\rvert^{2} + \lvert\textbf{h}_{{\rm L},m}^{H} \textbf{w}^{\rm c} \lvert^{2}+\sigma_{a}^2},
\end{aligned}
\label{MSINR}
\end{equation}
where  ${\textbf w}^{\rm c}_{n} \in \mathbb{C}^{N_{\rm T} \times 1}$ and ${\textbf w}^{\rm p}_{k} \in \mathbb{C}^{N_{\rm T} \times 1}$ denote the LEO satellite beamforming vectors associated with the common signal stream and private signal stream, respectively. Next, the corresponding common SINR and private  SINR at the $k$-th LGU are, respectively, given by
\begin{equation}
\begin{aligned}
\Gamma_{k}^{\rm c}\left(\{\textbf{w}^{\rm p}_{k}, \textbf{w}^{\rm c}\} \right)= \frac{ \lvert \textbf{h}_{{\rm L},k}^{H}\textbf{w}^{\rm c}\rvert ^{2}}{	 \sum\limits_{k'=1}^{K}\lvert \textbf{h}_{{\rm L},k}^{H}\textbf{w}_{n,k'}^{\rm p}\rvert^{2} 
+\sum\limits_{m=1}^{M}\lvert \textbf{h}_{{\rm G},k}^{H}\textbf{w}_{m}\rvert^{2}
+ \sigma_{b}^2},
\end{aligned}
\label{FSINR}
\end{equation}
and
\begin{equation}
\begin{aligned}
\Gamma_{k}^{\rm p}\left(\{\textbf{w}^{\rm p}_{k}\}\right)= \frac{ \lvert \textbf{h}_{{\rm L},k}^{H}\textbf{w}^{\rm p}_{k}\rvert ^{2}}{	 \sum\limits_{k'\neq k}^{K}\lvert \textbf{h}_{{\rm L},k}^{H}\textbf{w}^{\rm p}_{k'}\rvert^{2} 
+\sum\limits_{m=1}^{M}\lvert \textbf{h}_{{\rm G},k}^{H}\textbf{w}_{m}\rvert^{2}+ \sigma_{b}^2}.
\end{aligned}
\label{FSINR_v1}
\end{equation}

\subsubsection{Channel Models} 
Channel modeling in satellite communication models is quite different, such as fixed and time-varying channels \cite{9575181}. For the fixed channel model, it assumes that the {channel statistical characteristics} remain unchanged over a long time, where the corresponding channel model is expressed by
\begin{equation}
\label{Eq_channelgain}
\textbf{h}=\sqrt{G_{s}G_{k}(\frac{c}{4\pi f_{c}d_{s}})^{2}}\textbf{g},
\end{equation}
where $G_{s}$ and $G_{k}$ denote the satellite antenna gain and the user antenna gain, respectively. $c$ is the light speed, $f_c$ is the carrier frequency, and $d_s$ is the distance between the corresponding satellite and the user. $\textbf{g}$ is the small-scale fading vector with Rician distribution.  For notational simplification, we omit the index of $\textbf{h}_{{\rm L}, k}$ and $\textbf{h}_{{\rm G},m}$ here.

For the time-varying channel model, it is typically that {channel statistical characteristics} such as Doppler shift change rapidly with time. Following Jakes model \cite{nasir2019multi}, keeping the large-scale fading in (\ref{Eq_channelgain}) unchanged, the small-scale fading vector $\textbf{g}$ is  modeled as a first-order complex Gauss Markov process, i.e.,
\begin{equation}
\label{Eq_smallscale}
{\textbf{g}}(t)= \rho {\textbf{g}}{(t-1)} + \sqrt{1-\rho^{2}}\textbf{e}.
\end{equation}
In  (\ref{Eq_smallscale}), $\textbf{e}$ denotes the additive complex Gaussian noise with the same distribution as $\textbf{g}$ and the correlation coefficient $\rho$ is determined by
\begin{equation}
\rho = J_{0}(2\pi f_{d} T_{s}),
\end{equation}
where $J_{0}(\cdot)$, $T_{s}$, and $f_{d}$ denote the first kind zero-order Bessel function, the time interval between adjacent instants, and the maximum Doppler frequency, respectively.


\subsubsection{Optimization Goals}
The optimization goals of satellite communications focus on improving various performance indicators, such as maximizing SE and EE \cite{ardah2019hybrid}.

For SE,  the achievable information rate at  the $m$-th GGU is given by
\begin{equation}
R_{m}\left(\{\textbf{w}^{\rm p}_{k}, \textbf{w}^{\rm c}\} \right)={\rm log}_2(1+\Gamma_{m}{\left(\{\textbf{w}^{\rm p}_{k}, \textbf{w}^{\rm c}\} \right)}).
\end{equation}
Similarly, the achievable information rates of  the common signal part and the private part at the $k$-th LGU are, respectively, given by
\begin{equation}
R^{\rm c}_{k}\left(\{\textbf{w}^{\rm p}_{k}, \textbf{w}^{\rm c}\} \right)={\rm log}_2(1+\Gamma^{\rm c}_{k}\left(\{\textbf{w}^{\rm p}_{k}, \textbf{w}^{\rm c}\} \right)),
\end{equation}
and 
\begin{equation}
R^{\rm p}_{k}\left(\{\textbf{w}^{\rm p}_{k}\} \right)={\rm log}_2(1+\Gamma^{\rm p}_{k}\left(\{\textbf{w}^{\rm p}_{k}\} \right)).
\end{equation}
To guarantee that the common signal is successfully decoded at all  LGUs, the rate of the common message should be chosen as ${\min_k} R^{\rm c}_{k}\left(\{\textbf{w}^{\rm p}_{k}, \textbf{w}^{\rm c}\} \right)$.
Denote $c_{k}$ as the data rate for receiving the common message at the $k$-th LGU, which should satisfy that
\begin{flalign}
\sum\limits_{k = 1}^K {c_{k}} \leq  \mathop {\min }\limits_k R^{\rm c}_{k}\left(\{\textbf{w}^{\rm p}_{k}, \textbf{w}^{\rm c}\} \right).
\end{flalign}
Then, the sum achievable rate of the LEO area is expressed as
\begin{flalign}
&R\left(\{\textbf{w}^{\rm p}_{k}, \textbf{w}^{\rm c}, c_{k}\}\right) = \sum\limits_{k = 1}^K \left({c_k}+  R^{\rm p}_{k}\left(\{\textbf{w}^{\rm p}_{k}\} \right)\right).
\end{flalign}

For EE, it is a ratio of the sum achievable rate to the total power consumption and the system EE at the LEO area is expressed by
\begin{equation}
{\rm EE}\left(\{\textbf{w}^{\rm p}_{k}, \textbf{w}^{\rm c}, c_{k}\}\right)=\frac{R\left(\{\textbf{w}^{\rm p}_{k}, \textbf{w}^{\rm c}, c_{k}\}\right)}{P_{\rm T}\left(\{\textbf{w}^{\rm p}_{k}, \textbf{w}^{\rm c}\}\right)},
\end{equation}
where 
\begin{equation}
P_{\rm T}\left(\{\textbf{w}^{\rm p}_{k}, \textbf{w}^{\rm c}\}\right)= \mu \left(\left\|\textbf{w}^{\rm c}\right\|^{2}  + \sum\limits_{k=1}^{K}\left\|\textbf{w}_{k}^{\rm p}\right\|^{2}\right) + P_{\rm C},
\end{equation}
with $\mu \in [1,\infty)$ and $P_{\rm C}$ being the power amplifier efficiency factor and the constant power consumption by circuit modules, respectively.

\begin{remark}
It can be observed that human users need to carefully configure the above four aspects to build a reasonable satellite model manually. Nonetheless, errors are prone to occur during model building due to the varying situations and multiple choices. For example, if the user neglects rate ordering when choosing RSMA as the access protocol or mistakenly adopts stable channels to communicate with rapidly moving satellites, the model correctness and the corresponding communication efficiency might be significantly affected.
\end{remark} 
 
\begin{figure*}[tpb]
\centering
\includegraphics[width=\textwidth]{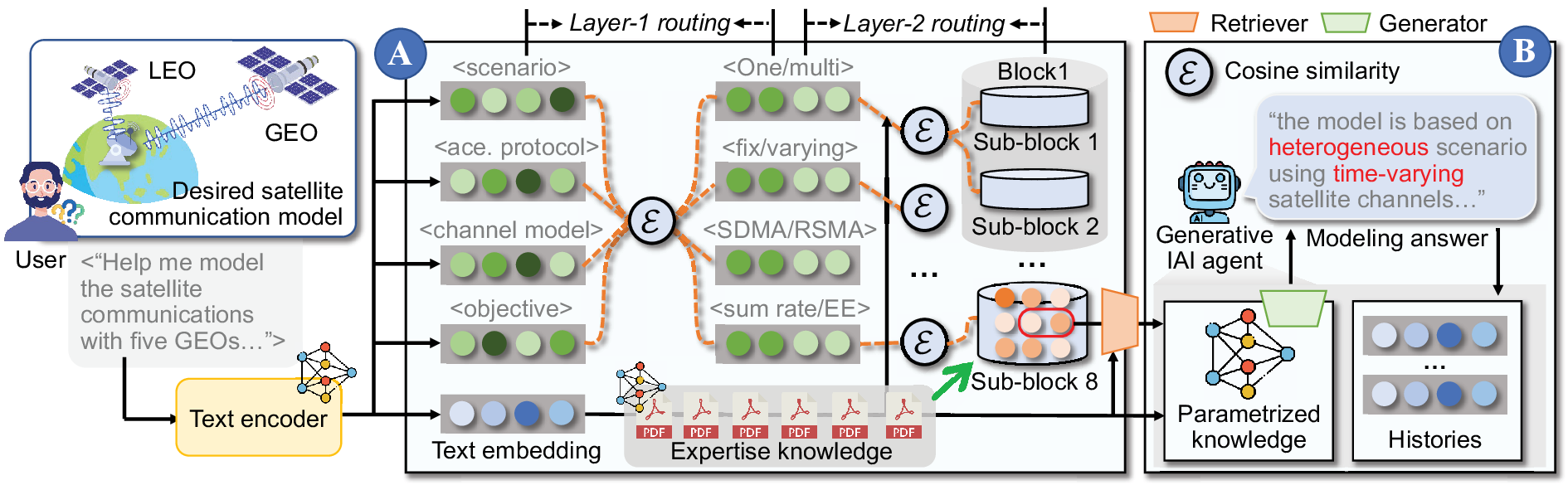}
\caption{The design of generative AI agent for satellite communication modeling. \textbf{A}: The two-layer semantic router. The expertise knowledge is organized into 4 blocks, each of which contains 2 sub-blocks. \textbf{B}: The answer generation for satellite communications modeling.}
\label{Architecture}
\end{figure*}

To this end, a generative AI agent framework is constructed, as shown in Fig. \ref{Architecture}. Specifically, we first utilize LLMs to establish a conversational modeling procedure.
In each round of interaction, the semantic router \cite{SemanticRouter} integrated into LLM extracts task-relevant semantics from users' natural language and invokes the corresponding function calls. 
Since LLMs are trained on general knowledge and lack satellite expertise, we build a RAG \cite{lewis2020retrieval} system, which contains massive expert knowledge regarding satellite communications to support complicated mathematical modeling.
The RAG functions called by the semantic router retrieve relevant knowledge from the dataset.
Afterward, the agent leverages LLMs to analyze the retrieved knowledge and generate the configurations of each modeling aspect step-by-step, based on user descriptions and requirements.
Next, we introduce the detailed design.

\subsection{Semantic Router}
{As shown in Fig.~\ref{Architecture}, we segment the entire expertise dataset into four blocks, each corresponding to one aspect mentioned in Section III-A. 
Moreover, each block contains two sub-blocks, corresponding to the specific configurations in each aspect. 
In this case, to generate accurate system models, the prerequisite for our generative agent is to locate the correct sub-block for expertise retrieval. 
The semantic router acts as the nerves of agents, transforming users' descriptions in natural language to the calling for the specific sub-block. 
To do so, it employs a two-layer structure, in which layer-1 routing realizes the mapping between user description and block.
Afterward, the layer-2 routing further determines which sub-block to call. 
The two routing selection layers are based on semantic similarity \cite{SemanticRouter}. 
Specifically, denoting a user description/query as $q \in \mathbf{q}$, the semantic router will use an encoder $\mathbf{E}_q$ to encode it as text embeddings, i.e., $\mathbf{E}_q(q)$. 
Here, text embeddings play a crucial role in bridging humans and LLMs by encoding natural language into a numerical format that neural networks can process.  In this paper, we leverage the pre-trained \textit{text-embedding-ada-002} \cite{Enbedding} model to generate embeddings of textual inputs. By training on large corpora, this model acquires outstanding contextual understanding capabilities, projecting each word/phrase into the continuous feature space with minimized conflicts. Moreover, we adopt the widely used \textit{cosine distance} \cite{SemanticRouter} to measure the semantic similarity between two embeddings.} 

The labels of blocks and sub-blocks are $\mathbf{b}_i$ and ($\mathbf{b}_i$, $\mathbf{s}^i_j$), where $\mathbf{b}_i$ and $\mathbf{s}^i_j$ ($i \in \{1, 2, 3, 4\}, j \in \{1, 2\}$) refer to the aspect and configuration, respectively. For instance, $\mathbf{b}_3$ = ``\textit{channel model}'', $\mathbf{s}^3_1$ = ``\textit{SDMA}'', and $\mathbf{s}^3_2$ = ``\textit{RSMA}'' (see Fig. \ref{Architecture}). Using cosine similarity as the metric for similarity measurement, the layer-1 routing of query $q$ can be defined as follows:
\begin{equation}
    \min_{i} \left(1.0 - \frac{\sum\limits_{k=1}^{\zeta}\left(\mathbf{E}_q(q)[k]\times\mathbf{E}_q(\mathbf{b}_i)[k]\right)}{\sqrt{\sum\limits_{k=1}^{\zeta}\left(\mathbf{E}_q(q)[k]\right)^2}\sqrt{\sum\limits_{k=1}^{\zeta}\left(\mathbf{E}_q(\mathbf{b}_i)[k]\right)^2}} \right),
    \label{19}
\end{equation}
where $\zeta$ denotes the length of text embeddings.
From this equation, we can observe that the user query will be routed to the block whose label is the most relevant.
Similarly, layer-2 routing repeats Eq. (19) while aiming to further route $q$ to the most relevant sub-block.

\begin{remark}
The semantic router is presented to enhance the efficiency of RAG for satellite communications modeling. This is because the rate of correct retrieval decreases with the increasing knowledge volume. By organizing expertise knowledge into eight sub-blocks, the space for retrieval can be reduced by eight times, making our proposal suitable for complicated satellite communication modeling. Moreover, we note that the block and sub-block configuration is customizable. Users can include/remove blocks/sub-blocks according to the specific application scenarios. After routing the user query to the specific sub-block, the RAG-assisted modeling can be performed, which is discussed in the next part.
\end{remark}

\subsection{Retrieval-Augmented Generation}
RAG aims to retrieve relevant expertise from the knowledge sub-block to support satellite communications modeling.
Such a process can be modeled as using input sequence \(\query\) to fetch knowledge \(\doc\), subsequently utilized as supplementary context for helping LLMs to generate the modeling answers \(\answer\).
Inspired by \cite{lewis2020retrieval}, we develop our RAG system atop \emph{RAG-Sequence}, utilizing identical knowledge to forecast each token within the target sequence. 
As illustrated in Fig. \ref{Architecture}, our system encompasses two pivotal elements: (i) a retrieval mechanism \(p_\retparam(\doc|\query)\) characterized by parameters \(\retparam\), tasked with deriving a truncated distribution over knowledge embeddings in response to a query \(\query\), and (ii) a generative module \(p_\genparam(\answer_i|\query, \doc, \answer_{1:i-1})\), parameterized by \(\genparam\), responsible for the sequential generation of tokens, each contingent upon the antecedent \(i-1\) tokens \(\answer_{1:i-1}\), the initial query \(\query\), and retrieved knowledge \(\doc\).
Such a process encompasses the marginalization of the seq2seq probability $p_{\rm Seq}(\answer|x)$ via a top-K approximation \cite{lewis2020retrieval}, formalized as follows:
\begin{equation}
\begin{aligned}
p_{\rm Seq}(\answer|q) &\approx \sum_{z \in {\rm top-K}(p(\cdot|q))} p_{\eta}(z|q)p(\answer|z, q) \\
&= \sum_{z \in top-K(p(\cdot|q))} p_{\eta}(z|q) \prod_{i}^{L} p_{\theta}(\answer_{i}| q,z, \answer_{1:i-1}),
\end{aligned}
\end{equation}
where $L$ represents the length of the sequence.
Note that top-K approximation is utilized to bring diversity to model output.
Next, we introduce the retriever and generator design, as well as the training and inference paradigms.

\begin{algorithm}[h]
    \caption{\textcolor{black}{{Generative AI agent for satellite communications modeling}}} \label{alg:5}
    $\mathbf{Input:}$ User query, embedding models, LLM;\\
    Initialize environment;\\
    Listen to user query $\mathbf{q} = \{q_1, q_2, \dots, q_n\}$\\
     \For{each $q$ in $\mathbf{q}$}{  
        Encode $q$ and acquire embeddings $\qenc(\query)$;\\
        \For{each $i$ in \{1, 2, 3, 4\}}{
            Calculate cosine similarity between $\qenc(\query)$ and $\qenc(\mathbf{b}_j)$ based on (\ref{19});\\
            }
        Route to the corresponding block $i$; \textit{\# layer-1 routing}\\
        \For{each $j$ in \{1, 2\}}{
            Calculate cosine similarity between $\qenc(\query)$ and $\qenc(\mathbf{s}^i_j)$ based on (\ref{19});\\
        }
        Route to the corresponding sub-block $j$; \textit{\# layer-2 routing}\\
        \For{each chunk in the sub-block}{
            Calculate cosine similarity between $\qenc(\query)$ and the chunk based on (\ref{19});\\
        }
        Fetch the most prevalent knowledge chunks;\\
        Call LLM to generate answers for $q$ based on retrieved knowledge;\\
    }
    $\mathbf{Output:}$ The entire satellite communications model containing scenario, access protocol, channel, and optimization goal;
    \end{algorithm}

\subsubsection{Retrieval Mechanism} The retriever function \(p_\retparam(\doc|\query)\) is predicated on the dense passage retriever (DPR) model\cite{lewis2020retrieval}, embracing a bi-encoder architecture, i.e.,
\begin{equation}
\begin{aligned}
p_\retparam(\doc|\query) \propto \exp \left(\denc(\doc)^{\top} \qenc(\query)\right),
\end{aligned}
\end{equation}
where $\denc(\doc)$ is a dense representation of the knowledge produced by the aforementioned text encoder \cite{kasneci2023chatgpt}, and $\qenc(\query)$ denotes the query representation produced by a query encoder. Due to its strong representation ability, such encoders are now usually implemented on top of LLM, such as text-embedding-ada-002 \cite{Enbedding}, i.e.,
\begin{equation}
\denc(\doc) = \text{LLM}_{e}(z)\;\; \& \;\;\qenc(\query)=\text{LLM}_{q}(\query).
\end{equation}
The knowledge in the selected sub-block is split into multiple chunks.
Based on Eq. (21), the most relevant chunks can be retrieved to support the generation of satellite communications modeling.

\subsubsection{Generative Module}
The generative module is pluggable and can be served by any mainstream LLM. Here, we adopt a generative pre-trained transformer (GPT) to realize \(p_\genparam(\answer_i|\query, \doc, \answer_{1:i-1})\), organizing retrieved knowledge and generating coherent and context-aware answers about satellite communications modeling. 

\subsubsection{Training Approaches}
The training scheme jointly optimizes the retriever and generator without explicit directives on document retrieval. 
The objective minimizes the negative marginal log-likelihood of the target sequences across a corpus of input/output pairs \((\query_j,\answer_j)\), employing stochastic gradient descent \textit{Adam}. 

\subsubsection{Generation Strategies}
During inference, \emph{RAG-Sequence} models adopt distinct approximations to $p_{\rm Seq}(\answer|q)$. 
This paradigm requires searching on each document \(\doc\), merging hypotheses across documents to form a candidate set.
The likelihood $p_{\rm Seq}(\answer|q)$ is then deduced by aggregating generator probabilities \(p_\genparam(\answer_i|\query, \doc, \answer_{1:i-1})\) across the document set, thereby generating rational and accurate answers about satellite communications modeling according to user descriptions/queries.
In summary, Algorithm \ref{alg:5} illustrates the process of performing customized satellite communications modeling via the proposed generative AI agent.

\subsection{Problem Formulation}

Using such a framework, human users can effectively tailor optimization problems to meet their unique requirements and objectives. For instance, when selecting a heterogeneous scenario with RSMA in a time-varying channel and focusing on SE, the human users aim to enhance the sum rate in LEO satellite regions. This involves a complex joint optimization of transmit private beamforming vectors $\{\textbf{w}^{\rm p}_{k}\}$, common beamforming vectors $\{\textbf{w}^{\rm c}\}$, and common rates $\{c_{k}\}$, constrained by the achievable rate requirements at GGUs and LGUs. 
Note that the sum rate maximization in time-varying channels for heterogeneous satellite networks with RSMA has remained an open issue thus far, where generative AI agent can help construct the formulated problem, i.e.,
\begin{subequations}
\label{ref_obj}
\begin{flalign}
\mathop{\max}\limits_{\{\textbf{w}^{\rm p}_{k}, \textbf{w}^{\rm c}, c_{k}\}}{\kern 1pt} {\kern 1pt} & {R} \left(\{\textbf{w}^{\rm p}_{k}, \textbf{w}^{\rm c}, c_{k}\}\right) \label{ref_a1} \\
{\rm s.t.}~~~&  \left\|\textbf{w}^{\rm c}\right\|^{2}  + \sum\limits_{k=1}^{K}\left\|\textbf{w}_{n}^{\rm p}\right\|^{2} \leq P_{\rm max}, \label{ref_b}\\
& R_{m}\left(\{\textbf{w}^{\rm p}_{k}, \textbf{w}^{\rm c}\} \right)\geq \xi_{\rm GGU}, \label{ref_c1}\\
&R_{k}\left(\{\textbf{w}^{\rm p}_{k}, \textbf{w}^{\rm c}, c_{k}\} \right) \geq \xi_{\rm LGU},\label{ref_d1} \\
&\sum\limits_{k = 1}^K {c_{k}} \leq  \mathop {\min }\limits_k R^{\rm c}_{k}\left(\{\textbf{w}^{\rm p}_{k}, \textbf{w}^{\rm c}\} \right) \label{ref_e1}\\
& c_{k} \geq 0,  \label{ref_f1} \\
& \forall m \in \mathcal{M}, \forall n \in \mathcal{N}, \forall k \in \mathcal{K}. \label{ref_g}
\end{flalign}
\end{subequations}
In constraint ($\rm \ref{ref_b}$), $P_{\rm max}$ denotes the  maximum power budget at  LEO satellite.  In constraints ($\rm{\ref{ref_c1}}$) and ($\rm{\ref{ref_d1}}$), $\xi_{\rm GGU}$ and $\xi_{\rm LGU}$ denote the minimum required achievable information rate threshold of each GGU and LGU, respectively.  Constraints  ($\rm \ref{ref_e1}$) and ($\rm \ref{ref_f1}$) guarantee that the common message can be successfully decoded by each LGU. 

\begin{remark}
It is worth noting that Problem (\ref{ref_obj}) is just one of many potential configurations that a generative AI agent can customize. The framework is universally adaptable and can incorporate various human user-defined needs based on the database from above four aspects, such as non-orthgonal multiple access (NOMA) in protocols and reconfigurable intelligent surface (RIS) in scenarios. This adaptability highlights the utility of generative AI agents in building complex models that can be time-consuming to model and highly error-prone if relying solely on human expertise.
\end{remark}

\begin{proposition}
For any how customized modeling is adopted based on the above four aspects (i.e., scenarios, access protocols, channel models, and optimization goals), the formulated problem is an NP-hard problem.
\begin{proof}
To prove the NP-hardness of formulated problem, it is equivalent to reducing the formulated problem to one of the proven NP-hard problems.  Particularly, for any how customized modeling is adopted, with given some optimization variables such as $\left\{{c_k} \right\}$ or $\textbf{w}^{\rm c}$, the formulated problem is reduced to a multi-ratio fractional programming problem, which has been reported to be an NP-hard problem in \cite{shen2018fractional}.
\end{proof}
\end{proposition}

\section{Proposed PPO with MoE Approach} \label{PPO}

{As demonstrated in Proposition 1, due to the existence of heterogeneous variables and the complexity of the optimization environment, Problem (\ref{ref_obj}) is an NP-hard problem, which poses a significant challenge in finding reasonable solutions.} Therefore, in this section, we introduce an MoE-PPO approach. {Note that generative AI agent can generate a variety of optimization problems, we focus on addressing one of the most challenging formulated problems, i.e., (\ref{ref_obj}), constructed by the generative AI agent. For other simpler formulated problems that the proposed AI agent constructs, fewer experts and a more straightforward action space can be used.}

\subsection{Overview of MoE}

\subsubsection{Architecture of the MoE Model}
MoE is the integration of multiple specialized sub-networks (i.e., ``experts'') under the guidance of a central ``gating network'' \cite{ye2023taskexpert}. This design consists of multiple expert networks ($E_1, E_2, ..., E_I$), each with unique parameters and running simultaneously. The central gating network labeled $G_{\sigma}$ generates a weight vector that enables the model to direct attention to the most relevant experts based on the current input.

In the MoE architecture, each expert is usually designed as a feed-forward neural network, processing inputs independently and producing outputs with the same dimensions, where the core function of the MoE model is described as
\begin{equation}
\label{MoE_1}
y = \sum\limits_{i=1}^{I}G_{\sigma}(x)_{i}E_{i}(x),
\end{equation}
In (\ref{MoE_1}), $E_i(x)$ is the output from the $i$-th expert network for a given input. The gating network $G_{\sigma}(x)$ typically utilizes a Softmax function, i.e.,
\begin{equation}
G_{\sigma}(x) = \frac{1}{1+{\rm exp}({-x \cdot W_{g})}},
\end{equation}
where $W_g$ is a learnable weight matrix. This design ensures that the MOE model leverages the most pertinent experts for any specific input through $G_{\sigma}(x)$'s selective activation mechanism, thereby optimizing the model's performance. 

\subsubsection{MoE Model Training}
For the training stage, MoE involves adjusting the parameters of the expert network and the gating network to minimize the loss function that reflects the difference between the predicted output and the real output, where the loss function based on  mean-squared error (MSE) is defined as
\begin{equation}
\mathcal{L} = \frac{1}{B} \sum_{b=1}^{B}\left(y_b - \sum\limits_{i=1}^{I}G_{\sigma}(x_{b})_{i}E_{i}(x_{b})\right)^2,
\end{equation}
where $B$ represents the total number of training samples, $y_b$ denotes the actual output for the $b$-th sample, and $x_b$ denotes the corresponding input. Next, we introduce an MOE-PPO framework, showing how to integrate task decomposition and policy optimization to further improve the learning performance of PPO.


\subsection{Mixture-of-Experts with PPO (MoE-PPO) Framework}

Integrating the MoE with the PPO approach, we enhance the policy's capacity to model complex behaviors in environments. Specifically, the MoE-PPO framework adopts an actor-critic structure, where the actor network, parameterized by ${\bm \theta}_{\rm A}$, encapsulates a dynamic ensemble of expert policies for decision-making. Concurrently, the critic network, parameterized by ${\bm \theta}_{\rm C}$, assesses the state-value function, $V_{{\bm \theta}_{\rm C}}(\mathtt{s})$, guiding the actor's policy improvement.

\subsubsection{Actor network with MoE}
The actor network in the MoE-PPO framework is defined as
\begin{equation}
    \pi_{{\bm \theta}_{\rm A}}(a_t|\mathtt{s}_t) = \sum_{i=1}^{I} w_{\theta_{i}}(\mathtt{s}_t) \pi_{\psi_i}(a_t|\mathtt{s}_t),
\end{equation}
where each $\pi_{\psi_i}$ represents an expert policy within the mixture, and $w_{\theta_{i}}(\mathtt{s}_t)$ denotes the dynamic weighting of expert $i$ for state $\mathtt{s}_t$, managed by the routing function parameterized by $\theta$.

\subsubsection{Gating Function with PPO}
The gating function in the MOE model plays a crucial role by dynamically selecting and aggregating expert policies based on the current state $\mathtt{s}_t$. These selecting and aggregating processes, however, involves making decisions from a categorical distribution, which presents challenges for optimization within the PPO framework. To tackle this issue, a new approach to estimate the gradient of the actor network under gating function is introduced.

Given a state $\mathtt{s}_t$, the value of advantage $A_{\phi}(\mathtt{s}_t, a_t^i)$ decides the contribution of each expert policy $\pi_{\psi_i}$ to final action decisions. For each available action $a_t^i$ sampled from the respective expert policy $\pi_{\psi_i}(\cdot|\mathtt{s}_t)$, we compute the advantage $A_{\phi}(\mathtt{s}_t, a_t^i)$, reflecting the expected return of selecting $a_t^i$ in state $\mathtt{s}_t$. The optimal expert policy for $\mathtt{s}_t$ is identified as the one yielding the highest advantage.

\begin{definition}[Back propagation maximization]
For optimizing actor network, the gradient related to the expert $i$, denoted as $\text{grad}_{\theta_{A_{i}}}$, is estimated using the back propagation maximization approach, i.e., 
\begin{equation}
\text{grad}_{\theta_{A_{i}}} = \delta_{\theta_{A_{i}}} \nabla_{\theta_{A_{i}}} \pi_{\theta_{A_{i}}}(\mathtt{s}_t), 
\end{equation}
where 
\begin{equation}
\quad \delta_{\theta_{A_{i}}} = \mathbbm{1}_{\{i=\arg\max_j A_{\phi}(\mathtt{s}_t, a_t^j)\}} - \pi_{\theta_{A_{i}}}(\mathtt{s}_t),
\end{equation}
with $\mathbbm{1}_{\{i=\arg\max_j A_{\phi}(\mathtt{s}_t, a_t^j)\}}$ being an indicator function that equals 1 if expert $i$'s action $a_t^i$ maximizes the advantage $A_{\phi}(\mathtt{s}_t, a_t^j)$ over all experts, and 0 otherwise. 
\end{definition}



Based on the back propagation maximization, the MOE-PPO can be effectively optimized to refine the overall strategy $\pi_{{\bm \theta}_{\rm A }}$ from multiple experts' policies.

\subsubsection{Objective Function with MOE}
To align the MOE framework within the PPO, the surrogate objective function is adapted to accommodate the MoE while maintaining the core principles of PPO \cite{zhang2023energy}, which is expressed by
\begin{equation}
    J\left({\bm \theta}_{\rm A}\right) = \mathbb{E}_t\left[ \rho_t\left({\bm \theta}_{\rm A}\right) A_t^{\pi_{{\bm\theta}_{\rm A}^{\rm old}}}\right],
\end{equation}
where  $\rho_t\left({\bm \theta}_{\rm A}\right) = \frac{\pi_{{\bm \theta}_{\rm A}}(a_t|\mathtt{s}_t)}{\pi_{{\bm\theta}_{\rm A}^{\rm old}}(a_t|\mathtt{s}_t)}$ denotes the probability ratio of the chosen action under the current policy to the old policy, reflecting the essence of the importance sampling technique used in PPO to estimate the expected advantage of the policy update. $A_t$ is the advantage function, which is used for highlighting the role of each expert in the collective policy through their weighted contributions to the estimated rewards, i.e.,
\begin{equation}
    A_t^{\pi_{{\bm\theta}_{\rm A}^{\rm old}}} = r(\mathtt{s}_t, \mathtt{a}_t) + \gamma V_{{\bm\theta}_{\rm C}^{\rm old}}(\mathtt{s}_{t+1}) - V_{{\bm\theta}_{\rm C}^{\rm old}}(\mathtt{s}_t).
    \end{equation}

However, the monotonic improvement theory requires that ${\pi_{{\bm \theta}_{\rm A}}(\mathtt{a}_t|\mathtt{s}_t)}$ and ${\pi_{{\bm\theta}_{\rm A}^{\rm old}}(\mathtt{a}_t|\mathtt{s}_t)}$ satisfy the following trust region constraint, i.e., 
 \begin{flalign}\label{KL}
 \mathbb{E}_t\left[ {\rm KL} \left({\pi_{{\bm \theta}_{\rm A}}(\mathtt{a}_t|\mathtt{s}_t)}  || {\pi_{{\bm\theta}_{\rm A}^{\rm old}}(\mathtt{a}_t|\mathtt{s}_t)} \right) \right] \le \varepsilon,
 \end{flalign}
 where ${\rm KL}(\cdot)$ represents the Kullback-Leibler divergence function, and $\varepsilon$ is a positive hyperparameter.

To facilitate the computation, $J\left({\bm \theta}_{\rm A}\right)$ with ${\bm \theta}_{\rm A}$ satisfying (\ref{KL}) can be approximated by
\begin{flalign}\label{clip}
&J^{\rm CLIP}\left({\bm\theta}_{\rm A}\right)= \\
&\mathbb{E}\left[ {{\min}\left(\rho_t\left({\bm\theta}_{\rm A}\right)A_t^{\pi_{{\bm\theta}_{\rm A}^{\rm old}}},{\rm clip}\left(\rho_t\left({\bm\theta}_{\rm A}\right),1-\varepsilon, 1+\varepsilon\right)A_t^{\pi_{{\bm\theta}_{\rm A}^{\rm old}}}\right)} \right],\nonumber
\end{flalign}
where ${\rm clip(\cdot)}$ is the clip function to restrain $\rho_t({\bm\theta}_{\rm A})$  to lie in the range $[1-\varepsilon,1+\varepsilon]$ and $\varepsilon$ is a hyperparameter which is decayed during training stage.

\subsubsection{Policy and Critic Update Mechanisms}
\begin{figure}[!t]
\centering
\includegraphics[width=0.49\textwidth]{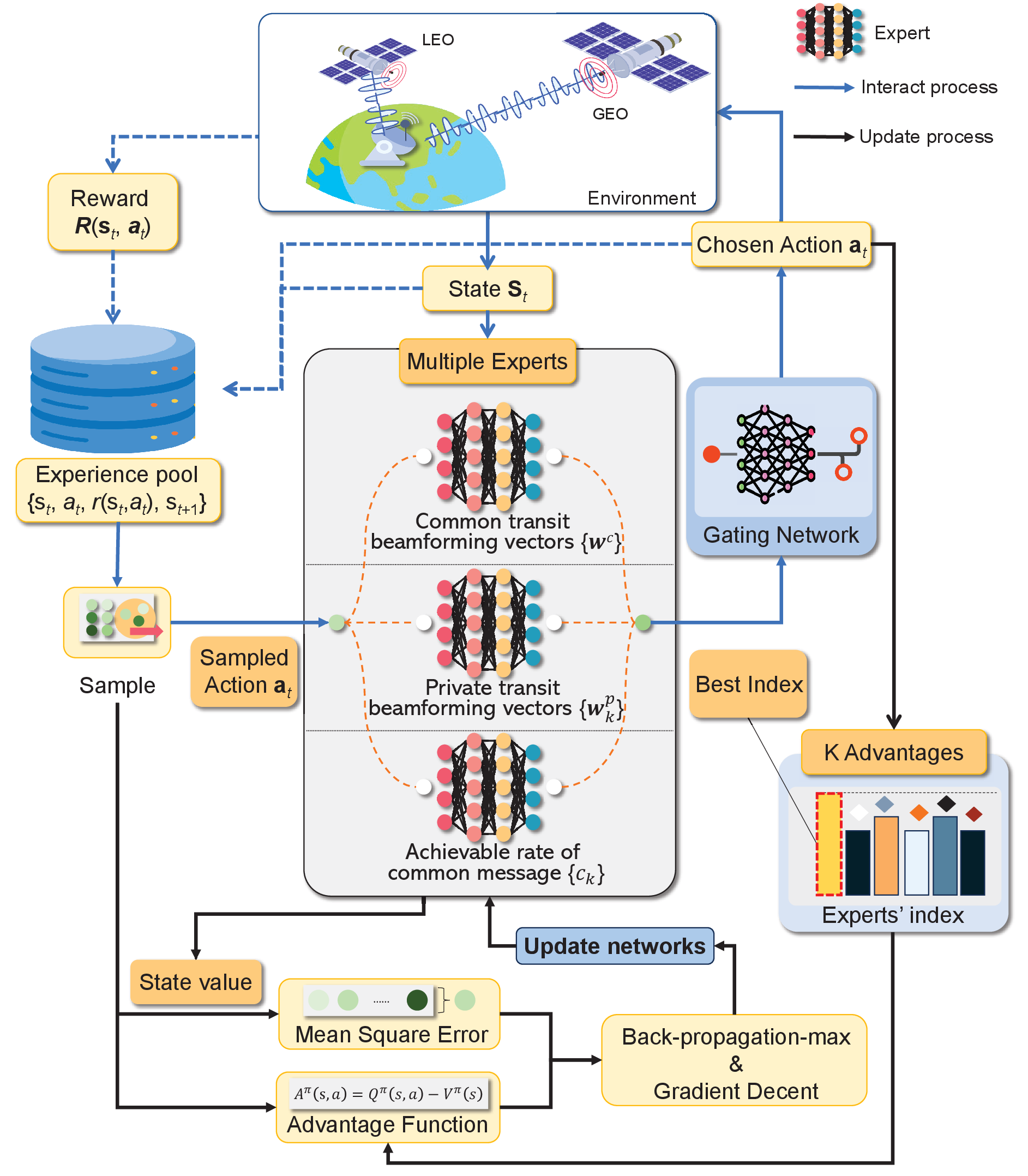}
\caption{The diagram of the proposed MoE-PPO approach.}
\label{RL_MoE}
\end{figure}

The parameters of both the actor and critic networks are refined through mini-batch stochastic gradient descent (SGD) using experiences sampled from the environment, where their update rules are, respectively, formulated as
\begin{align}
    {\bm\theta}_{\rm A} &= {\bm\theta}^{\rm old}_{\rm A} - \alpha_{\rm A}\frac{1}{B}\sum\nolimits_{t=1}^B\left({ \nabla_{{\bm\theta}_{\rm A}}{\tilde J}_t^{\rm CLIP}\left({\bm\theta}_{\rm A}\right) }\right),
    \label{Q_A}
\end{align}
and
\begin{align}
    {\bm\theta}_{\rm C} &= {\bm\theta}^{\rm old}_{\rm C} - \alpha_{\rm C}\frac{1}{B}\sum\limits_{t=1}^B \nabla_{{\bm\theta}_{\rm C}}\left(V_{{\bm\theta}_{\rm C}}\left(\mathtt{s}_{t}\right) - V_{{\rm tar}}\left(\mathtt{s}_{t}\right)\right)^2,
    \label{Q_C}
\end{align}
where ${\tilde J}_t^{\rm CLIP}\left({\bm\theta}_{\rm A}\right)$ reflects the realization of the CLIP-modified objective function, and $V_{{\rm tar}}\left(\mathtt{s}_{t}\right)$ represents the target value for the state $\mathtt{s}_t$, which is computed by
\begin{flalign}\label{td_v}
{V_{{\rm tar}}\left(\mathtt{s}_{t}\right)= r\left(\mathtt{s}_t,\mathtt{a}_t\right) + \cdots + \gamma^{n+1} V_{{\bm \theta}^{\rm old}_{\rm C}}\left(\mathtt{s}_{t+n+1}\right)},
\end{flalign}
where $\gamma \in [0,1]$ denotes a discount factor. For clarity, the diagram of the proposed MoE-PPO approach is shown in Fig.~\ref{RL_MoE}.

\begin{remark}
{Note that, in our proposed MoE-PPO approach, we use a back-propagation maximization method to update the experts. This method uses the gradient of the most advantageous function value across all experts to update each expert, effectively allowing the best-performing expert to guide the improvement of the other experts. Additionally, the gating network dynamically adjusts weights, ensuring that the influence of poorly performing experts is mitigated. By doing so, the impact of the worst-performing experts on the whole network is significantly minimized. Such the back-propagation maximization method and gating network have been adopted in existing studies, e.g., \cite{ren2021probabilistic} and \cite{obando2024mixtures}.}
\end{remark}

\subsection{MDP Formulation}

In MDP, we define the framework through the components $\langle \mathcal{S}, \mathcal{A}, \mathcal{R}, \gamma \rangle$. These components encapsulate the state space $\mathcal{S}$, action space $\mathcal{A}$, reward function $\mathcal{R}$, and a discount factor $\gamma$, where the detailed designs are shown as follows.


\subsubsection{Action Space}
In action space, each expert is assigned to generate specific actions in terms of $\{{\textbf{w}^{\rm P}_{k}}\}$,  $\textbf{w}^{\rm c}$,  and ${c_{k}}$. Considering the complexity of handling complex-valued beamforming vectors in neural networks and inspired by \cite{clerckx2019rate}, a decomposition approach is adopted, where each beamforming vector $\{{\textbf{w}}_{v}\}$ (including  $\{{\textbf{w}^{\rm P}_{k}}\}$ and ${\textbf{w}^{\rm c}}$ ) is split into its magnitude and a unit-norm direction vector, i.e., 
\begin{equation}
\textbf{w}_{v} = \left\|\textbf{w}_{v}\right\| \overline{\textbf{w}}_{v},  \forall v  \in \{1,2,...,K+1\}.
\label{splitt}
\end{equation}
In (\ref{splitt}), $\left\|\textbf{w}_{v}\right\|$ denotes the  magnitude of the corresponding transmit power and  $\overline{\textbf{w}}_{v}$  denotes the unit-norm beam direction vector.

For the magnitude of the corresponding transmit power  $\left\|\textbf{w}_{v}\right\|$, it is determined by the hyperbolic tangent function to ensure that the power constraints are met, which is expressed by 
\begin{equation}
\left\|\textbf{w}_{v}\right\| = \frac{\sqrt{P_{\rm max}}}{2} \left({\underbrace{ \frac{{\rm exp}(x^{\rm POW}_{v})-{\rm exp}(-x^{\rm POW}_{v})}{{\rm exp}(x^{\rm POW}_{v})+{\rm exp}(-x^{\rm POW}_{v})}}_{\rm hyperbolic \ tangent \ function}}+1\right),
\end{equation}
{where $\frac{{\rm exp}(x^{\rm POW}_{v})-{\rm exp}(-x^{\rm POW}_{v})}{{\rm exp}(x^{\rm POW}_{v})+{\rm exp}(-x^{\rm POW}_{v})} \in [-1,+1]$ is  employed as an activation function to guarantee that the outputs of the deep neural networks (DNNs), i.e., the selected transmit power, satisfy the corresponding
constraints  \cite{9110869}.} $x^{\rm POW}_{v} =f(W \cdot x^{\rm POW'}_{v}+b^{l})$  denotes the output of the activation function with respect to the corresponding selected transmit power. $f(\cdot)$, $W$,  $x^{\rm POW'}_{v}$, and $b^{l}$ denote the activation function of the current layer, the corresponding weight, the output in the previous layer and the corresponding bias, respectively.

For the unit-norm beam direction vector $\overline{\textbf{w}}_{v}$, following \cite{clerckx2019rate}, we leverage the maximum ratio transmission (MRT) and  zero forcing (ZF) strategies, i.e.,
\begin{flalign}
\overline{\textbf{w}}_{v}=
\begin{cases}
\frac{\sum\nolimits_{k = 1}^K{\bf{\tilde h}}_{{\rm L}, k}^{H}}{||\sum\nolimits_{k = 1}^K{\bf{\tilde h}}_{{\rm L}, k}^{H}||},  ~~{\rm if}~ v=K+1,  \\
\frac{\textbf{v}_{k}}{||\textbf{v}_{k}||},  ~~{\rm if}~ v \neq K+1,  \\
\end{cases}
\end{flalign}
where ${{\tilde{\bf h}}_{{\rm L},k}} = {{\bf{h}}^{H}_{{\rm L}, k}}$ and ${\bf v}_k$ is the $k$-th column of ${\bf{V}} = [{\bf{v}}_{1}, ..., {\bf{v}}_{K}]$ where $\textbf{V} = {\bf G}^{H}({\bf G}{\bf G}^{H})^{-1}$ and ${\bf G} = [{\bf{{\tilde h}}}_{{\rm L}, 1}, ..., {\bf{{\tilde h}}}_{{\rm L}, K} ]$.

For the achievable rate of the common message $\{{{c}_{k}}\}$, it is also determined by the hyperbolic tangent function to ensure that the common rate constraints are met, which is expressed by 
\begin{flalign}
\label{EQ_c_k}
c_{k} =   &\frac{\mathop {\min }\limits_k R^{\rm c}_{k}\left(\{\textbf{w}^{\rm p}_{k}, \textbf{w}^{\rm c}\} \right)}{2}\left( \underbrace{\frac{{\rm exp}(x^{\rm COM}_{k})-{\rm exp}(-x^{\rm COM}_{k})}{{\rm exp}(x^{\rm COM}_{k})+{\rm exp}(-x^{\rm COM}_{k})}}_{\rm hyperbolic \ tangent \ function}+1\right),
\end{flalign}
where $x^{\rm COM}_{k}$  denotes the output of the activation function with respect to the corresponding selected achievable rate of the common message.

Therefore, the action space is defined as
\begin{equation}
\label{actions}
\mathcal{A}= \left\{ \left\{{\textbf{w}^{\rm p}_{k}}\right\},  {\textbf{w}^{\rm c}}, \left\{{{c}_{k}}\right\} \right\},
\end{equation}
where the cardinal number of the action space $\mathcal{A}$ is $(2K+1)$.

\subsubsection{State Space}

The action space comprises the set of current information of all the users $\textbf{u}  \in \mathbb{R}^{(2K+M)\times 1} $, the selected action vector $\textbf{a}$, and the instant reward $r$.

For the  current information of all users $\textbf{u}$, it is expressed by
\begin{flalign}
\label{stete}
\textbf{u} = \left\{\left\{{\Gamma}^{\left( {\rm{C}} \right)}_{k}\right\}, \left\{{\Gamma}^{\left( {\rm{P}} \right)}_{k}\right\},  \left\{{\Gamma}_{m}\right\} \right\},
\end{flalign}
where ${\Gamma}^{\left( {\rm{C}} \right)}_{k}$, ${\Gamma}^{\left( {\rm{P}} \right)}_{k}$, and ${\Gamma}_{m}$ represent the corresponding SINR of GGU and LGU in the last time step.

For the selected action vector $\textbf{a}$, it is chosen from the set of possible actions ${\cal A}$, which reflects the strategy implemented in the last time step. The instant reward $r$ is generated from the state and the chosen action, which indicates the efficacy with which the agent addresses the described problem. In the subsequent subsection, we provide a definition of instant reward $r$  relevant to our proposed approach.

Therefore, the state space is defined as
\begin{flalign}
\label{state}
\mathcal{S} = \left\{ \textbf{u}, \textbf{a}, r \right\},
\end{flalign}
where the cardinal number of the state space $\mathcal{S}$ is {$(4K+M+1)$}.

\subsubsection{Reward Function}
The reward function considers both the objective function and the constraints of the Problem, which comprises two elements: the direct reward term, which reflects the unconstrained energy efficiency (EE), and the penalty term, which ensures compliance with the constraints. To harmonize the relationship between the direct reward and the penalty term, we define a penalty-based reward function that balances these two aspects, i.e.,
\begin{align}
\label{reward}
r= &{\rm{R}}\left(\{\textbf{w}^{\rm p}_{k}, \textbf{w}^{\rm c}, c_{k}\}\right)\\
& \times \underbrace{(\Omega_{\rm Pow} \times \Omega_{\rm Com}\times \Omega_{\rm LEO} \times \Omega_{\rm GEO})}_{\rm{Penalty \ term }},\nonumber
\end{align}
where
\begin{flalign}
\Omega_{\rm Pow} =
\begin{cases}
1,  ~~ \left\|\textbf{w}^{\rm c}\right\|^{2}  + \sum\limits_{k=1}^{K}\left\|\textbf{w}_{k}^{\rm p}\right\|^{2} \leq  {P_{\rm max}}, ~ \forall k \in \mathcal{K},  \\
0,  ~~\left\|\textbf{w}^{\rm c}\right\|^{2}  + \sum\limits_{k=1}^{K}\left\|\textbf{w}_{k}^{\rm p}\right\|^{2}  > {P_{\rm max}},  ~ \forall k \in \mathcal{K},\\
\end{cases}
\end{flalign}
\begin{flalign}
\Omega_{\rm Com} =
\begin{cases}
1,  ~~ \sum\nolimits_{k = 1}^K {c_k} \leq \mathop {\min }\limits_k R^{\rm c}_{k}\left(\{\textbf{w}^{\rm p}_{k}, \textbf{w}^{\rm c}\} \right), ~ \forall k \in \mathcal{K},  \\
0,  ~~\sum\nolimits_{k = 1}^K {c_k} >  \mathop {\min }\limits_k R^{\rm c}_{k}\left(\{\textbf{w}^{\rm p}_{k}, \textbf{w}^{\rm c}\} \right), ~ \forall k \in \mathcal{K}, \\
\end{cases}
\end{flalign}
\begin{flalign}
\Omega_{\rm LEO} =
\begin{cases}
1,  ~~\mathop {\min }\limits_k R_{k}\left(\{\textbf{w}^{\rm p}_{k}, \textbf{w}^{\rm c}, c_{k}\} \right) \geq \xi_{\rm LGU},~\forall k \in \mathcal{K},  \\
0,  ~~\mathop {\min }\limits_k R_{k}\left(\{\textbf{w}^{\rm p}_{k}, \textbf{w}^{\rm c}, c_{k}\} \right)  < \xi_{\rm LGU}, ~\forall k \in \mathcal{K}, \\
\end{cases}
\end{flalign}

and
\begin{flalign}
\Omega_{\rm GEO} =
\begin{cases}
1,  ~~\mathop {\min }\limits_m (R_{m}\left(\{\textbf{w}^{\rm p}_{k}, \textbf{w}^{\rm c}\} \right))  \ge \xi_{\rm GEO}, ~\forall m \in \mathcal{M},  \\
0,  ~~\mathop {\min }\limits_m (R_{m}\left(\{\textbf{w}^{\rm p}_{k}, \textbf{w}^{\rm c}\} \right))  < \xi_{\rm GEO}, ~\forall m \in \mathcal{M}. \\
\end{cases}
\end{flalign}
$\Omega_{\rm Pow}$, $\Omega_{\rm Com}$, $\Omega_{\rm LEO}$, and  $\Omega_{\rm GEO}$ respectively represent the penalties for actions without satisfying the power budget requirement (\ref{ref_b}), the common message decoding requirement (\ref{ref_e1}), the LEO QoS requirement (\ref{ref_d1}), and the GEO QoS requirement (\ref{ref_c1}). 

\begin{remark}
The penalty terms have been widely used to guarantee a feasible solution, which, however, may induce sparse rewards and high variance. Hence,  we only adopt the penalty for the complicated constraint, i.e., (\ref{ref_b})-(\ref{ref_e1}), while the simple constraint, i.e., (\ref{ref_f1}) is guaranteed via the activation function based on (\ref{EQ_c_k}).
\end{remark}

Note that a non-zero reward is awarded only when all the constraints of Problem $\rm P_0$ are met, which means that our objective can be achieved. 
Therefore, for the agent to receive positive rewards, it must optimize the sum rate and ensure that all constraints are fully satisfied. 
In summary, Algorithm \ref{alg:3} describes the training procedure for the proposed MoE-PPO approach.
 \begin{algorithm}[h]
    \caption{\textcolor{black}{{The proposed MoE-PPO based approach}}} \label{alg:3}
    $\mathbf{Input:}$ Episodes, exploration steps, update episodes, learning rate, channel status information, number of experts;\\
    Initialize environment;\\
    Observe initial environment state $\mathbf{s}_0$
     \For{each itertion}{    
        \For{ each exploration step $t$}{
            Each expert makes its own action $a^{'}_{t}$ based on sub-policies;\\
            Aggregate actions $a_t= \{\textbf{w}^{\rm p}_{k}, \textbf{w}^{\rm c}, c_{k}\}$ based on the current state $s_t$ and gating mechanism;\\
            Obtain the current reward $r_t$;\\
            Observe the next state $s_{t+1}$;\\
            Store transition $ \left( s_t, a_t, r_t, s_{t+1}\right)$ in $\mathcal{D}$;
            }
        \For{each update step t}{
            Sample $B$ mini-batch of transitions from $\mathcal{D}$ from all experts;\\
            Update actor network $\theta_{A}$ and critic network $\theta_{C}$ via SGD according to (\ref{Q_A}) and (\ref{Q_C}) for half episodes of total update episodes;\\
            Calculate advantage function $A(s,a)$ for all experts and obtain the maximization value $A^{'}(s,a)$ through Back-propagation maximization approach; \\
            Update actor network $\theta_{A}$ and critic network $\theta_{C}$ via SGD and $A^{'}(s,a)$ according to (\ref{Q_A}) and (\ref{Q_C}) for half episodes of total update episodes;\\
            Update state $s_t \gets  s_{t+1}$;
        }
    }
    $\mathbf{Output:}$ The actions $\{\textbf{w}^{\rm p}_{k}, \textbf{w}^{\rm c}, c_{k}\}$;
    \end{algorithm}


\section{Simulation Results} \label{sim}
This section presents simulations to evaluate our proposed generative AI agent framework and our MoE-PPO approach. 

\subsection{Simulation Parameters and Setup}

\subsubsection{Scenario Settings}

Since our customized system is based on heterogeneous scenarios, the simulated scenario is composed of one LEO and one GEO.   The number of antennas at GGU and LGU are both set as $8$. The number of GGU  and LGU are both set to 2. The power budget of GEO and  LEO are both set to 50 $\rm{dBm}$. Following \cite{huang2023deep}, the achievable information rate requirements of GGU and LGU are set as 1 $ {\rm b/s/Hz}$ and  0.1 $ {\rm b/s/Hz}$, respectively. 
The noise power of each GGU and LGU are both set as -104 $\rm{dBm/Hz}$.  Moreover, the remaining hyper-parameters are summarized in Table~\ref{tab_PS}. 
\begin{table}[htbp]
\begin{small}
\begin{center}
\caption{Hyper Parameters Settings}
\label{tab_PS}
\centering
\begin{tabular}{|c|c|}  
\hline
\textbf{System Parameter}  &\textbf{Value} \\ \hline \hline %
Carrier frequency $f_{c}$ & 4 GHZ \\ \hline
Satellite antenna gain  $G_{s}$ & 35 dBi \\ \hline
LEO altitude  & 300 KM \\ \hline
GEO altitude & 4000  KM \\ \hline
Doppler frequency $f_{d}$ & 10Hz \\ \hline
Time interval $T_{s}$ & $2\times10^{-3}$s  \\ \hline
Rician factor & 4  \\ \hline
The number of NN layers  & 4   \\ \hline
Clipping parameter $\epsilon$  & 0.2  \\ \hline
DNN optimizer &  \emph{Adam}  \\ \hline
Activation function  &  ReLU and Tanh \\ \hline
Hidden layer size  &  256 \\ \hline
Learning rate $\alpha$  &  $3 \times 10^{-4}$ \\ \hline
Discount factor$\gamma$     &0 \\ \hline
Clipped probability ratio $\mu$  &0.2 \\ \hline
Memory size  & 40960 \\ \hline
Time slots & 10  \\ \hline
Batch size  &2048   \\ \hline
\end{tabular}
\end{center}
\end{small}
\end{table}

\begin{figure*}[tpb]
\centering
\includegraphics[width=\textwidth]{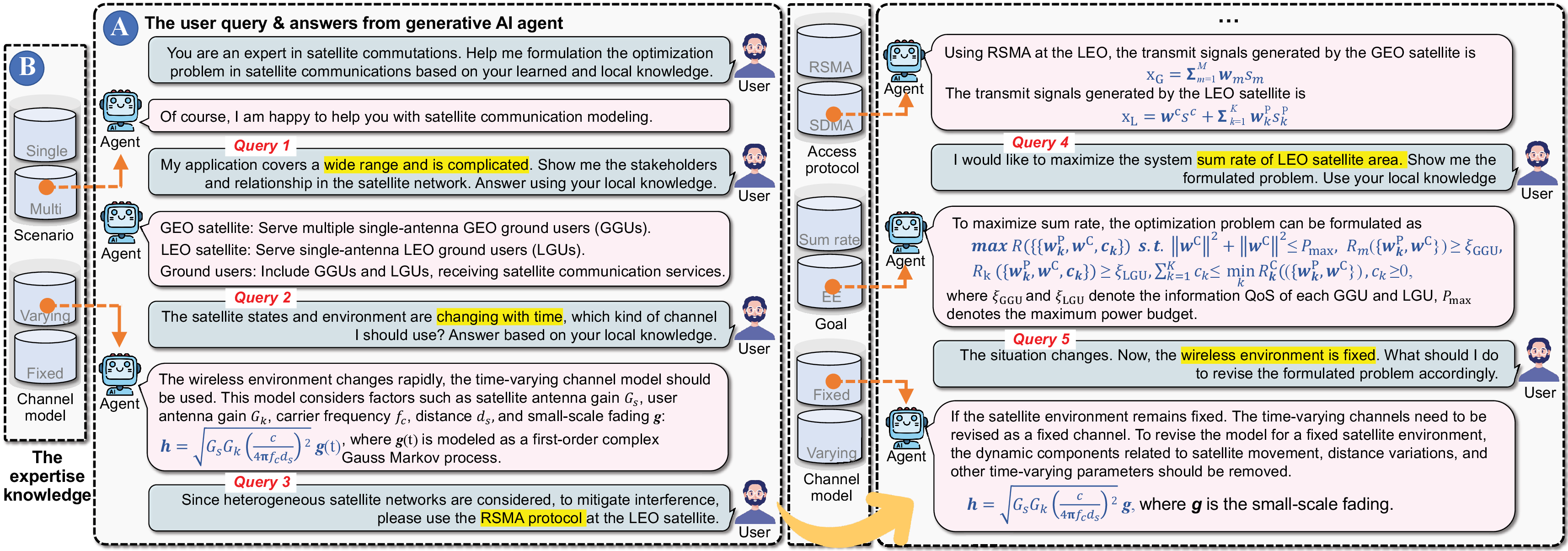}
\vspace{-0.6cm}
\caption{The process of satellite communication modeling using proposed generative AI agent. \textbf{(A)}: The user description and answers of generative AI agent. The words with key semantics are highlighted in yellow. \textbf{(B)}: The retrieval from expertise knowledge.}
\label{log}
\end{figure*}    

\subsubsection{Generative AI Agent Settings} 
The generative AI agent is implemented as follows. 
First, the encoders for user description and expertise knowledge are realized by \textit{OpenAIEmbeddings} \cite{Enbedding}.
The LLM for analyzing retrieved knowledge and generating satellite communication models is GPT-3.5 \cite{GPT-3.5}.
Finally, the semantic router, agent memory, and conversation chain are developed atop LangChain libraries \cite{Langchain}.

\subsubsection{MoE-PPO Approach Settings} For the proposed MoE-PPO approach, the whole structure is set as a four-layer feed-forward deep neural network, i.e., one input layer, one output layer, and two hidden layers. The total number of input ports and output ports are {$(4K +M +1)$} and $(2K + 1)$, respectively. Moreover, the remaining hyper-parameters are also summarized in Table~\ref{tab_PS}.
    
\begin{figure}[!t]
\centering
\includegraphics[width=0.49\textwidth]{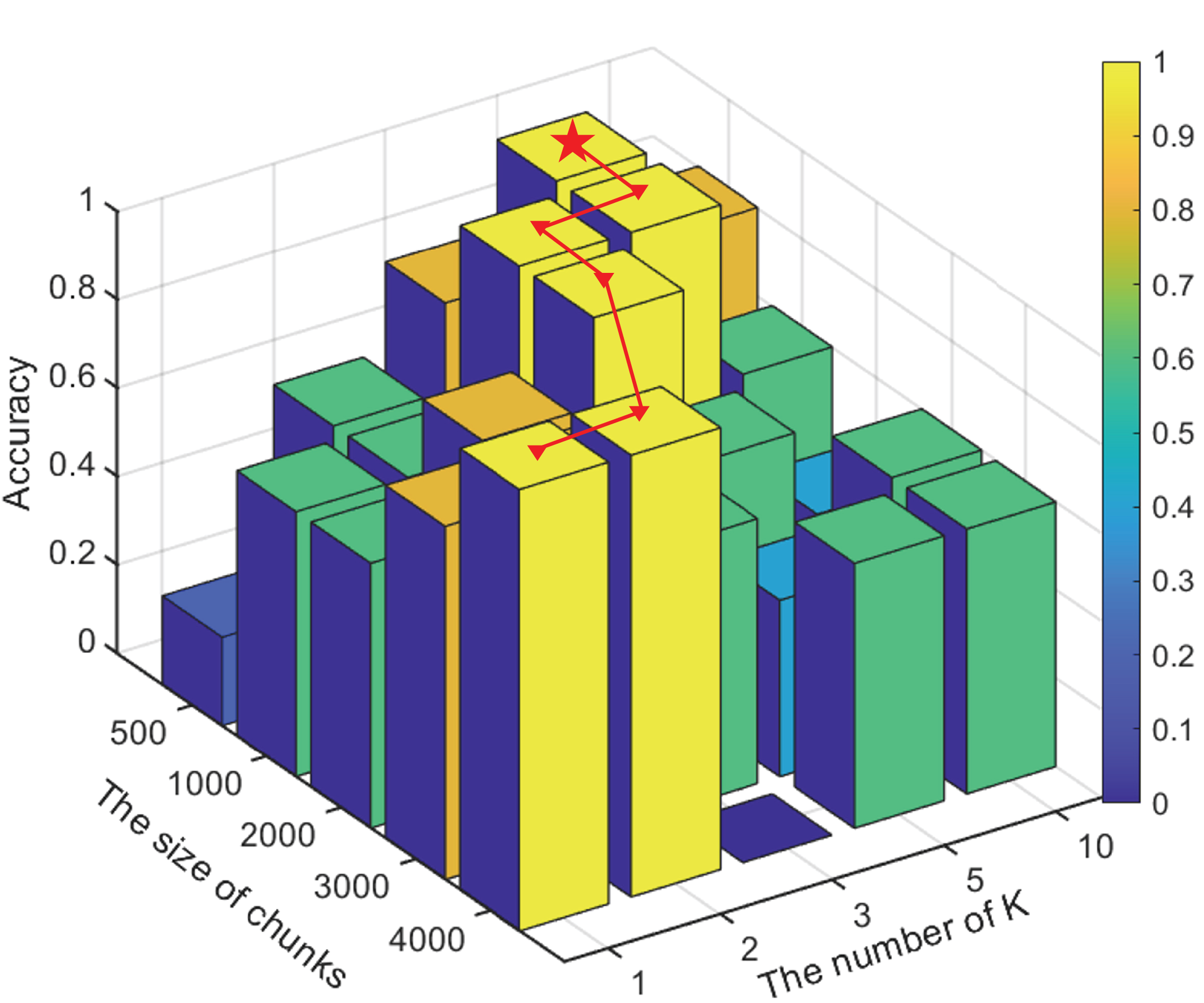}
\caption{The impact of chunk size and number on generation accuracy.}
\label{Chunk}
\end{figure}

\subsection{The Effectiveness of Generative AI Agent}
\subsubsection{Satellite Communications Modeling}
First, we evaluate the effectiveness of the proposed generative AI agent in customizing satellite communication models.
As shown in Fig. \ref{log}, the users apply the \textit{role assignment} strategy at the beginning, enabling the LLM to recall pre-trained satellite knowledge by asking it to act as satellite experts.
Then, the users describe the requirements for the satellite communication model from four aspects.
We can observe that our agent can precisely capture critical semantic keys from user natural language descriptions and call the corresponding RAG database.
Then, the LLM can leverage the retrieved knowledge to generate coherent answers about satellite communications modeling according to user requests.
Note that even though the user query is vague, the semantic router can still find the most relevant sub-block by semantic similarity.
For instance, the agent precisely routes to the time-varying channel without explicit mention by users.
In this case of Fig. \ref{log}, the user's requirement corresponds to a multiple-satellite network using RSMA protocol and time-varying channels, with the goal of maximizing the system EE.
We can observe that the AI agent successfully constructs the entire model using 6 rounds of interactions, reaching the minimum for accomplishing five-step modeling (one extra round is for \textit{role assignment}).
Moreover, with the varying network conditions, the user requirement may change over time.
Our agent can associate the requirement change to the parts of the models and revise such parts automatically, thus avoiding human-caused errors.
As illustrated in Fig.~\ref{log}, the agent precisely updates the channel model from Rayleigh fading to Rician when the network dynamics decrease.

\begin{figure}[!t]
\centering
\includegraphics[width=0.49\textwidth]{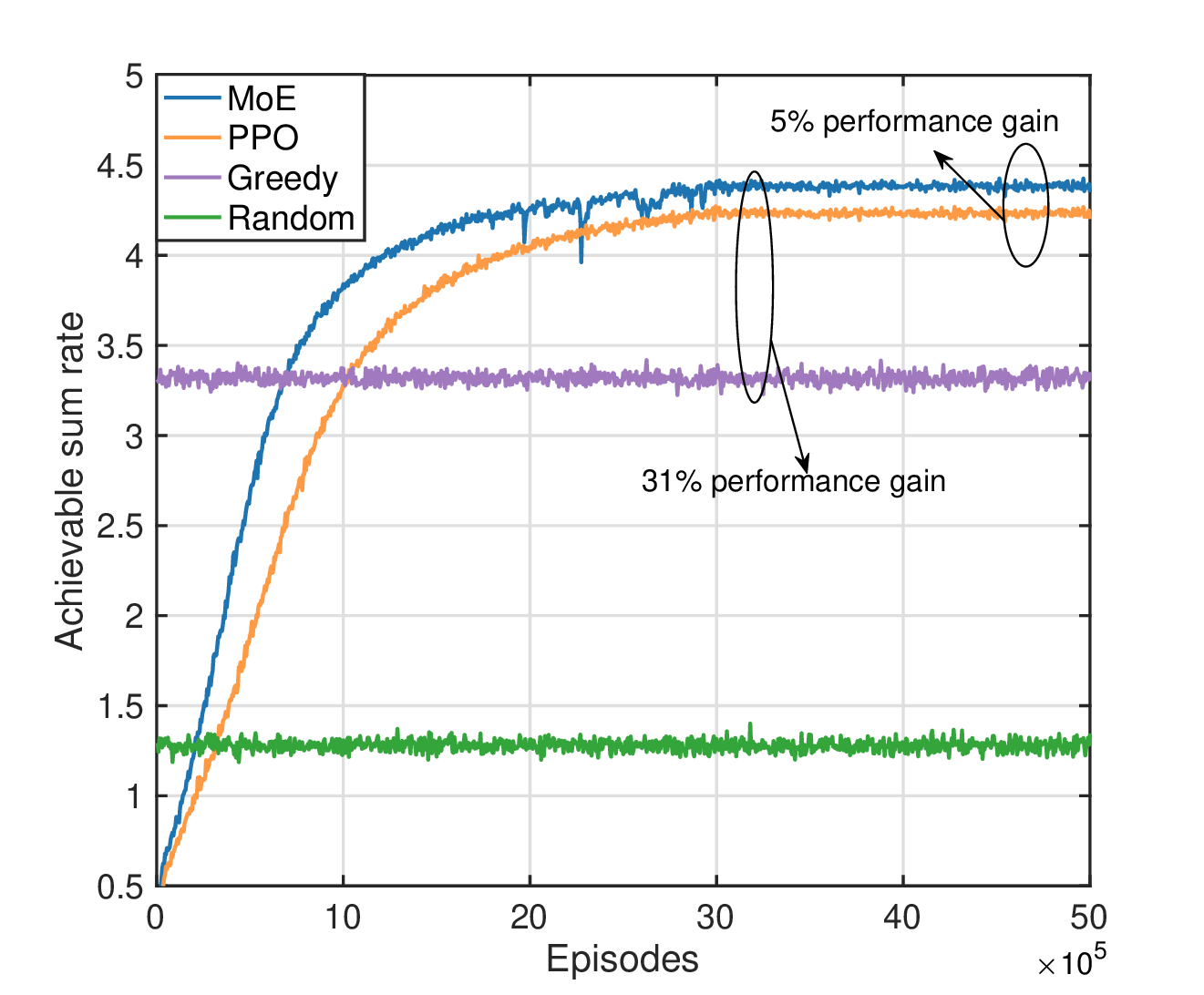}
\caption{The convergence of the proposed MoE-PPO approach.}
\label{Convergence}
\end{figure}

\subsubsection{Generation Accuracy}
{Next, we evaluate the accuracy of the generated model under different agent settings. For our proposed generative AI agent, it leverages RAG to help human users perform customized satellite communications modeling. Expert modeling knowledge is fed to the agent through chunked embeddings. If the correct knowledge chunk\footnote{Chunk is the unit of embedding segments. To increase the retrieval efficiency, in each sub-block, the agent splits the expert knowledge embedding into multiple chunks \cite{lewis2020retrieval}.} that corresponds to the specific user-described scenario is retrieved, the generated answer can be authentic and rational. Therefore, to check the authenticity and rationality of the generated content, the metric named retrieval rate (RR) \cite{chen2024benchmarking} can be utilized as accuracy. Specifically, RR indicates the ratio that the generative AI agent successfully retrieves the correct knowledge chunk and constructs the answer. Here, we adjust two hyperparameters in RAG, namely the chunk size and the number of chunks.
The former means the number of tokens in each chunk; the latter means the number of chunks that can be retrieved at one time.} Note that generation accuracy is defined as $n_c/5$, which means the number of correct answers among all five queries in Fig.~\ref{log}.
Fig.~\ref{Chunk} demonstrates that low chunk size and number lead to poor generation accuracy.
This is because the RAG system cannot fetch expertise to support answer generation.
However, if retrieving too much redundant expertise, the LLM of the agent can hardly analyze it effectively, resulting in poor accuracy as well.
Therefore, we set chunk size and number to be 500 and 5, respectively, which ensures that our agent can generate precise models with the minimum number of retrieved tokens (2500 in total).

\subsection{The Effectiveness of PPO with MoE Approach}

\begin{figure}[!t]
\centering
\includegraphics[width=0.49\textwidth]{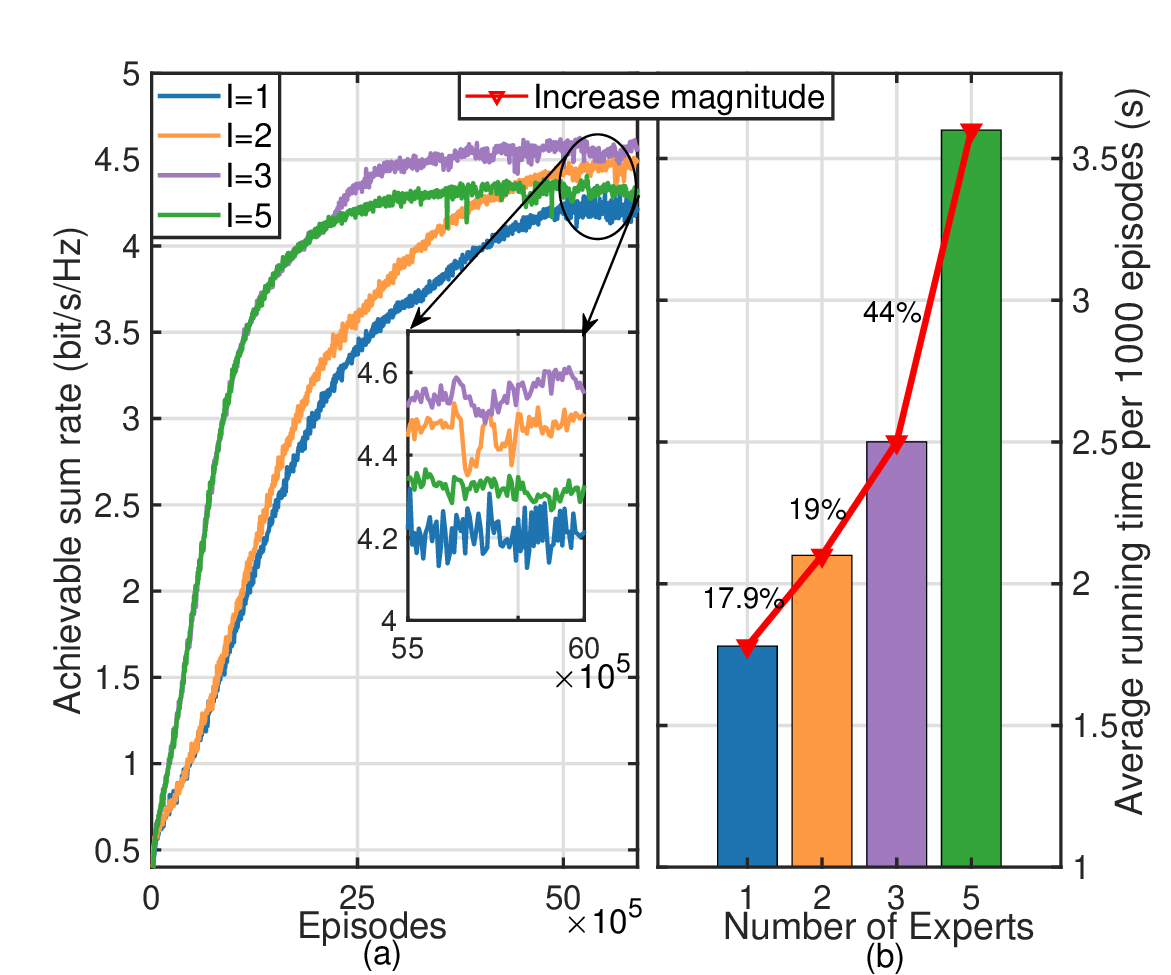}
\caption{The effectiveness of the different numbers of experts.}
\label{MoE_bijiao}
\end{figure}

\begin{figure}[!t]
\centering
\includegraphics[width=0.49\textwidth]{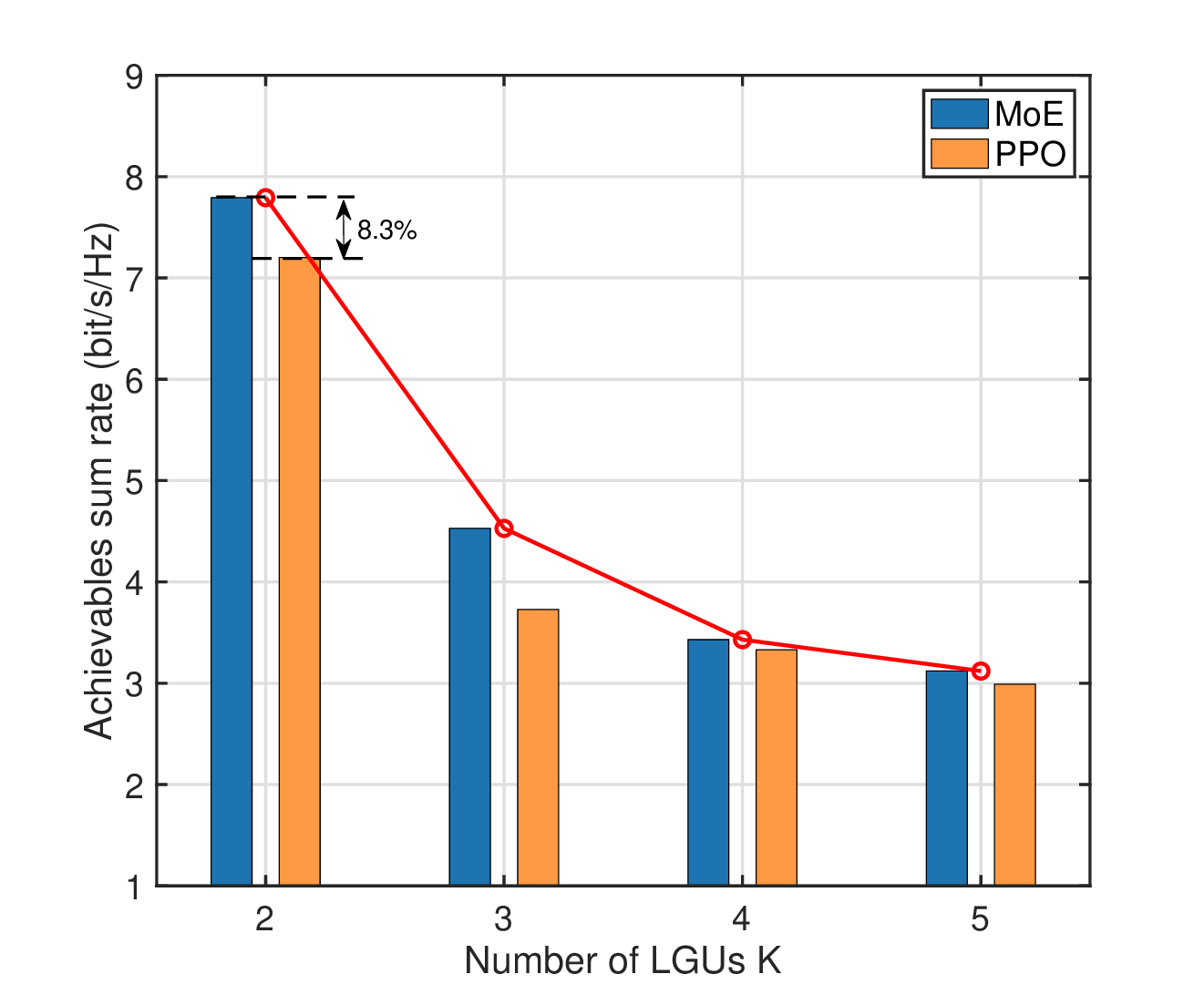}
\caption{\textcolor[rgb]{0,0,1}{The effectiveness of MoE-PPO under different number of LGUs.}}
\label{MoE_User}
\end{figure}

\subsubsection{The Convergence of PPO with MoE Approach}

Fig.~\ref{Convergence} illustrates the convergence behavior of the proposed MoE-PPO approach in comparison to some different benchmarks, such as the traditional PPO, greedy, and random selection strategies. The figure reveals that both the MoE-PPO and the standard PPO approaches demonstrate convergence as the number of episodes increases. It shows that MoE-PPO consistently surpasses the performance of PPO (i.e., about 5\%), achieving higher sum rates in each iteration. The performance enhancements of MoE-PPO can be attributed to the integration of the MoE framework, which utilizes expert policy specialization and adaptive policy weights to more effectively navigate the policy space. To further validate the effectiveness of MoE-PPO, its comparison with the greedy and random baselines after convergence underscores a significant improvement (i.e., about 220\% and 38\%). This improvement is facilitated by MoE-PPO's enhanced collaborative capabilities and more efficient utilization of the state-action space using gating network.

\begin{figure}[!t]
\centering
\includegraphics[width=0.49\textwidth]{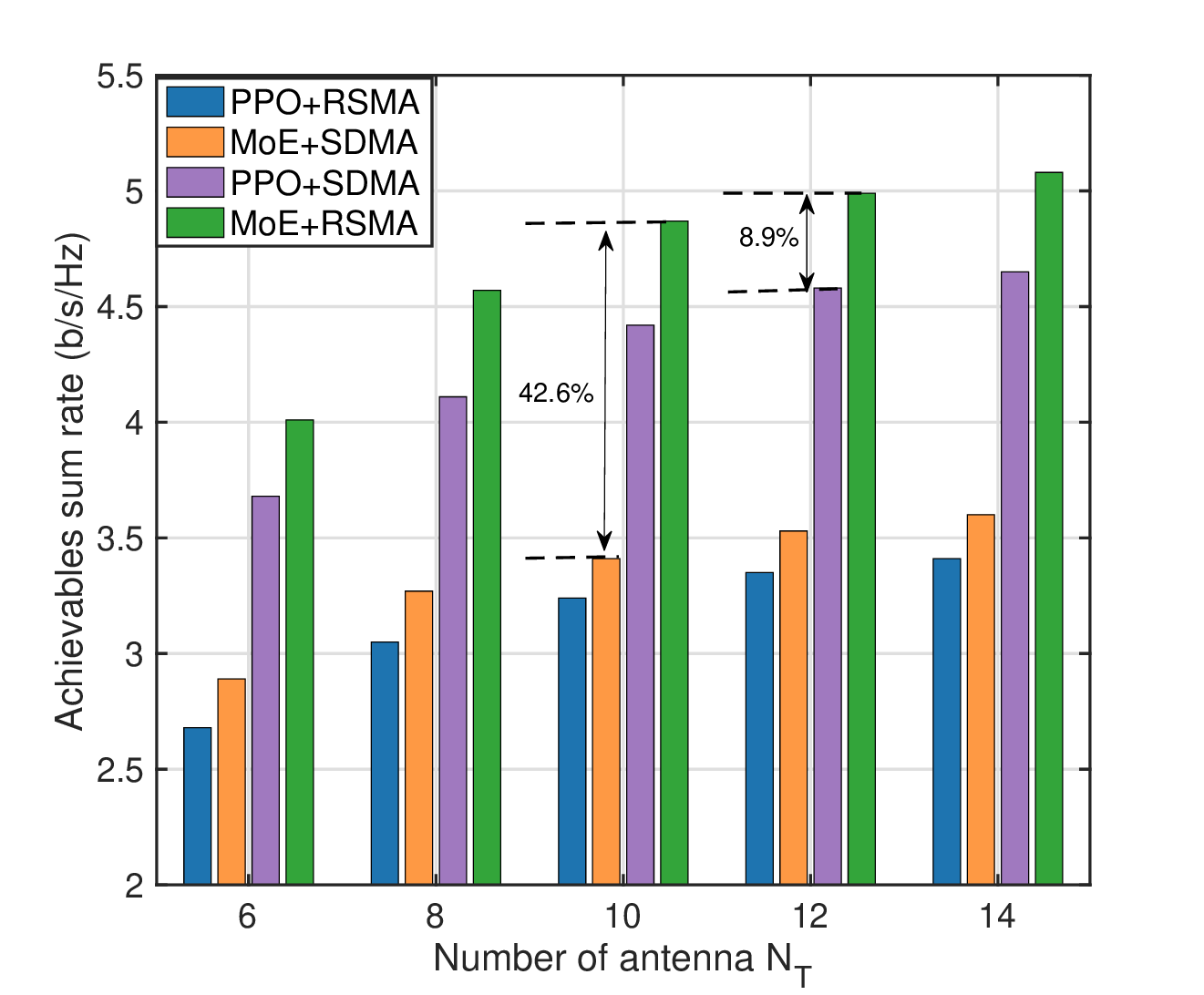}
\caption{SDMA design versus RSMA design by using different approaches.}
\label{SDMA}
\end{figure}
To validate the adaptability of the proposed MoE-PPO approach, we conducted studies under varying protocols (i.e., considering SDMA) and optimization goals (i.e., considering EE or power minimization).

\subsubsection{Different Number of Experts}
To illustrate the impact of the number of experts on optimization outcomes, we consider the scenarios with different numbers of experts:
\begin{itemize}
  \item  $I$=1: One expert optimizes all variables, i.e., $\{\textbf{w}^{\rm p}_{k}, \textbf{w}^{\rm c}, c_{k}\}$.
  \item $I$=2:  Two experts handle the tasks, with one expert optimizing the beamforming variables, i.e., $\{\textbf{w}^{\rm p}_{k}, \textbf{w}^{\rm c}\}$, and another expert focusing on the common rate variables $\{c_{k}\}$.
  \item $I$=3: Three experts handle the tasks, where the first expert is responsible for optimizing private beamforming variables, i.e., $\{\textbf{w}^{\rm p}_{k}\}$, the second is responsible for optimizing common beamforming variables, i.e., $\textbf{w}^{\rm c}$,  and the third expert is responsible for optimizing common rate variables, i.e., $\{c_{k}\}$.
  \item $I$=5: Five experts are involved, with each expert specializing in the optimization of a specific variable.
\end{itemize}
Fig.~\ref{MoE_bijiao} illustrates the impact of varying the number of experts within the MoE-PPO framework on system performance.  As depicted in Fig.~\ref{MoE_bijiao}(a), the MoE-PPO approach converges across the different numbers of experts, with the achievable sum rate improving as the number of experts increases. Notably, when the number of experts is set to three, the system achieves the highest sum rate. This optimal performance is attributed to the allocation of experts to distinct variable categories, enabling each expert to specialize in and optimize variables within their expertise domain, followed by an effective gating process that aggregates the optimized variables. In contrast, a decrease in performance is observed when the number of experts is five. This counterintuitive result can be explained by over-segmentation of the optimization task; each expert optimizes a single variable, leading to errors in the gating process due to lack of visibility into other variables or their corresponding category variables. Furthermore, Fig.~\ref{MoE_bijiao}(b) presents the average running time with the different numbers of experts. An exponential increase in running time is observed with more experts, a consequence of the additional computational overhead required to calculate and aggregate the decisions from a greater number of experts. Therefore, to balance effectiveness and efficiency, the number of experts should be equal to the types of variables to ensure that each expert can fully utilize their expertise without incurring excessive computational costs or gating errors.

\subsubsection{Different number of LGUs}
{{Fig.~\ref{MoE_User} illustrates the achievable sum rate as a function of the number of LGUs. It is observed that the achievable sum rate decreases as the number of users increases for both the proposed MoE-PPO approach and the pure PPO approach. This decline occurs because, with increasing users, the approach must allocate more communication resources to users with poor channel quality to meet constraints, reducing the overall sum rate. Moreover, it is evident that the proposed MoE-PPO method consistently outperforms the PPO method, where the MoE-PPO approach achieves an 8.3\% higher sum rate compared to the PPO approach when there are 2 LGUs. This superiority is due to the MoE's ability to leverage multiple specialized experts to handle different aspects of the optimization problem more effectively. Specifically, MoE-PPO can dynamically select the most relevant experts, optimizing resource allocation more efficiently than a single comprehensive network. }

\subsection{The Adaptability of the generative AI agent and MoE-PPO approach}

\subsubsection{MoE-PPO approach for different access protocols from generative AI agent construction}
Fig.~\ref{SDMA} presents the achievable sum rate obtained by RSMA and SDMA versus the number of transmit antennas $N_{t}$. It shows that the achievable sum rate increases with the number of $N_{t}$ for all considered protocols and approaches. This increment is attributable to the additional spatial degrees of freedom provided by more antennas, which help mitigate interference and enhance system capacity. Moreover, it is observed that the system performance under RSMA outperforms that under SDMA (about 42.6\% gain). This superiority of RSMA can be ascribed to the employment of successive interference cancellation (SIC) techniques, which effectively remove a part of the interference. Additionally, under each antenna configuration, our MoE-PPO approach always surpasses the PPO approach, which is consistent with the previous simulations.

\begin{figure}[!t]
\centering
\includegraphics[width=0.49\textwidth]{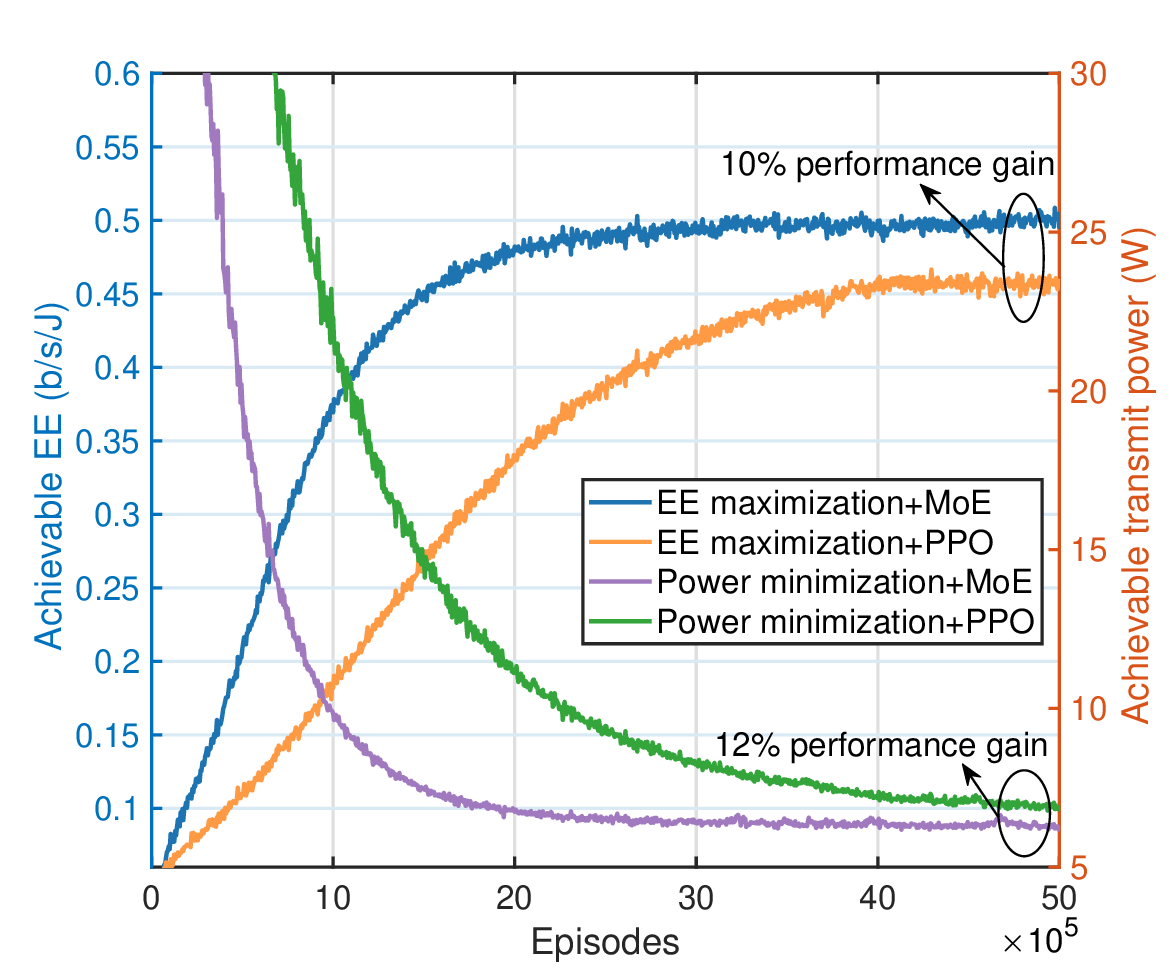}
\caption{The effectiveness of MoE-PPO under different optimization goals with $\mu=1$ and $P_{\rm C}=1$  .}
\label{MoE_EE}
\end{figure}

\subsubsection{MoE-PPO approach for different optimization goals from generative AI agent construction}
Furthermore, Fig.~\ref{MoE_EE} illustrates the different system performance of the proposed MoE-PPO approach under a time-varying mode with different optimization goals, i.e., EE maximization and transmit power minimization. In this context, the corresponding reward functions for MoE-PPO approach are respectively designed as follows.
\begin{itemize}
  \item  For EE maximization:
\begin{align}
r_{\rm EE}= &{\rm{EE}}\left(\{\textbf{w}^{\rm p}_{k}, \textbf{w}^{\rm c}, c_{k}\}\right) \\
&\times \underbrace{(\Omega_{\rm Pow} \times \Omega_{\rm Com}\times \Omega_{\rm LEO} \times \Omega_{\rm GEO})}_{\rm{Penalty \ term }}, \nonumber
\end{align}
  \item For power minimization: \begin{align}
r_{\rm P}= &\frac{\overbrace{(\Omega_{\rm Pow} \times \Omega_{\rm Com}\times \Omega_{\rm LEO} \times \Omega_{\rm GEO})}^{\rm{Penalty \ term }}}{{P_{\rm T}}\left(\{\textbf{w}^{\rm p}_{k}, \textbf{w}^{\rm c}\}\right)}.
\end{align}
\end{itemize}
It is observed that whether maximizing EE or minimizing power consumpution, our proposed MoE-PPO approach always outperforms the traditional PPO (about 10\% and 12\%) with the increment of episodes. The reason is that the MoE structure is able to efficiently allocate computing resources because each expert can operate in parallel to solve different variables in which they specialize. This parallelism facilitates deep exploration of the state space, resulting in better policy updates at each stage. Next, it also demonstrates the adaptability of the proposed MoE-PPO approach, confirming its ability to meet different channel conditions and optimization goals.

\section{Conclusion} \label{con}
This paper proposed a generative AI agent framework and an MoE-PPO method for network modeling and transmission strategy design in satellite communications networks, respectively. Specifically, the proposed framework utilized LLM and RAG for adaptive system modeling and configuration automation, while MoE-PPO optimized resource allocation and interference management by integrating expert knowledges. Simulation results confirmed the effectiveness of the generative AI agent framework and MoE-PPO approach in satellite communication networks.

\bibliographystyle{IEEEtran}
\bibliography{mylib}

\begin{thebibliography}{10}
\providecommand{\url}[1]{#1}
\csname url@samestyle\endcsname
\providecommand{\newblock}{\relax}
\providecommand{\bibinfo}[2]{#2}
\providecommand{\BIBentrySTDinterwordspacing}{\spaceskip=0pt\relax}
\providecommand{\BIBentryALTinterwordstretchfactor}{4}
\providecommand{\BIBentryALTinterwordspacing}{\spaceskip=\fontdimen2\font plus
\BIBentryALTinterwordstretchfactor\fontdimen3\font minus \fontdimen4\font\relax}
\providecommand{\BIBforeignlanguage}[2]{{%
\expandafter\ifx\csname l@#1\endcsname\relax
\typeout{** WARNING: IEEEtran.bst: No hyphenation pattern has been}%
\typeout{** loaded for the language `#1'. Using the pattern for}%
\typeout{** the default language instead.}%
\else
\language=\csname l@#1\endcsname
\fi
#2}}
\providecommand{\BIBdecl}{\relax}
\BIBdecl

\bibitem{8368236}
J.~Liu, Y.~Shi, and et. al., ``Space-air-ground integrated network: A survey,'' \emph{IEEE Commun. Surv. Tut.}, vol.~20, no.~4, pp. 2714--2741, 2018.

\bibitem{9502642}
S.~Liu, Z.~Gao, and et. al., ``L{EO} satellite constellations for {5G} and beyond: {H}ow will they reshape vertical domains?'' \emph{IEEE Commun. Mag.}, vol.~59, no.~7, pp. 30--36, 2021.

\bibitem{9749193}
H.~Chen, M.~Xiao, and Z.~Pang, ``Satellite-based computing networks with federated learning,'' \emph{IEEE Wireless Commun.}, vol.~29, no.~1, pp. 78--84, 2022.

\bibitem{9210567}
O.~Kodheli and et. al., ``Satellite communications in the new space era: {A} survey and future challenges,'' \emph{IEEE Commun. Surv. Tut.}, vol.~23, no.~1, pp. 70--109, 2021.

\bibitem{9403416}
L.~Qian, P.~Yang, Y.~L. Guan, and et. al., ``Multi-dimensional polarized modulation for land mobile satellite communications,'' \emph{IEEE Trans. Cogn. Commun. Netw.}, vol.~7, no.~2, pp. 383--397, 2021.

\bibitem{10209551}
P.~Yue, J.~An, J.~Zhang, and et. al., ``Low earth orbit satellite security and reliability: {I}ssues, solutions, and the road ahead,'' \emph{IEEE commun. Surv. Tut.}, vol.~25, no.~3, pp. 1604--1652, 2023.

\bibitem{345892}
A.~Jamalipour and et. al., ``Performance of an integrated voice/data system in nonuniform traffic low earth-orbit satellite communication systems,'' \emph{IEEE J. Sel. Areas Commun.}, vol.~13, no.~2, pp. 1--1, 1995.

\bibitem{10413484}
B.~Li and et. al., ``A novel frequency reuse model for co-existing {LEO} and {GEO} satellites,'' \emph{IEEE Wireless Commun. Lett.}, pp. 1--1, 2024.

\bibitem{9371230}
R.~Deng, B.~Di, H.~Zhang, and et. al., ``Ultra-dense {LEO} satellite constellations: {H}ow many {LEO} satellites do we need?'' \emph{IEEE Trans. Wireless Commun.}, vol.~20, no.~8, pp. 4843--4857, 2021.

\bibitem{9970355}
Z.~Xiao, J.~Yang, T.~Mao, and et. al., ``{LEO} satellite access network ({LEO-SAN}) towards {6G}: {C}hallenges and approaches,'' \emph{IEEE Wireless Commun.}, pp. 1--8, 2022.

\bibitem{park2023generative}
J.~S. Park and et. al., ``Generative agents: {I}nteractive simulacra of human behavior,'' in \emph{Proc. ACM USIT}, 2023, pp. 1--22.

\bibitem{zhang2024interactive}
R.~Zhang, H.~Du, Y.~Liu, D.~Niyato, and et. al., ``Interactive {AI} with retrieval-augmented generation for next generation networking,'' \emph{arXiv preprint arXiv:2401.11391}, 2024.

\bibitem{du2023user}
H.~Du, R.~Zhang, D.~Niyato, and et. al., ``User-centric interactive {AI} for distributed diffusion model-based {AI}-generated content,'' \emph{arXiv preprint arXiv:2311.11094}, 2023.

\bibitem{lewis2020retrieval}
P.~Lewis and et. al., ``Retrieval-augmented generation for knowledge-intensive {NLP} tasks,'' \emph{NeurIPS}, vol.~33, pp. 9459--9474, 2020.

\bibitem{zhang2023generative1}
R.~Zhang, H.~Du, and et. al., ``{G}enerative {AI} for space-air-ground integrated networks ({SAGIN}),'' \emph{arXiv preprint arXiv:2311.06523}, 2023.

\bibitem{6215056}
S.~E. Yuksel and et. al., ``Twenty years of mixture of experts,'' \emph{IEEE Trans. Neural Networks Learn. Syst.}, vol.~23, no.~8, pp. 1177--1193, 2012.

\bibitem{masoudnia2014mixture}
S.~Masoudnia and R.~Ebrahimpour, ``Mixture of experts: {a} literature survey,'' \emph{Artif. Intell. Rev.}, vol.~42, pp. 275--293, 2014.

\bibitem{chen2023mod}
Z.~Chen, Y.~Shen, M.~Ding, and et. al., ``Mod-{S}quad: {D}esigning mixtures of experts as modular multi-task learners,'' in \emph{Proc. IEEE/CVF CVPR}, 2023, pp. 11\,828--11\,837.

\bibitem{ACM}
J.~Zhang, Z.~Tang, M.~Li, and et. al., ``Crosssense: {T}owards cross-site and large-scale wifi sensing,'' in \emph{Proc. ICMCN}, 2018, pp. 305--320.

\bibitem{9419053}
Z.~Gao, A.~Liu, C.~Han, and X.~Liang, ``Sum rate maximization of massive {MIMO} {NOMA} in {LEO} satellite communication system,'' \emph{IEEE Wireless Commun. Lett.}, vol.~10, no.~8, pp. 1667--1671, 2021.

\bibitem{huang2023deep}
J.~Huang, Y.~Yang, J.~Lee, and et. al., ``Deep reinforcement learning based resource allocation for {RSMA} in {LEO} satellite-terrestrial networks,'' \emph{IEEE Trans. Commun.}, 2023.

\bibitem{khan2023rate}
W.~U. Khan, Z.~Ali, E.~Lagunas, and et. al., ``Rate splitting multiple access for next generation cognitive radio enabled leo satellite networks,'' \emph{IEEE Trans. Wireless Commun.}, 2023.

\bibitem{8957364}
J.~Li, K.~Xue, and et. al., ``Energy efficiency and traffic offloading optimization in integrated satellite/terrestrial radio access networks,'' \emph{IEEE Trans. Wireless Commun.}, vol.~19, no.~4, pp. 2367--2381, 2020.

\bibitem{du2024enhancing}
H.~Du, R.~Zhang, Y.~Liu, J.~Wang, Y.~Lin, Z.~Li, D.~Niyato, J.~Kang, Z.~Xiong, S.~Cui \emph{et~al.}, ``Enhancing deep reinforcement learning: A tutorial on generative diffusion models in network optimization,'' \emph{IEEE Commun. Surv. Tut.}, 2024.

\bibitem{wang2023does}
X.~Wang, Y.~Fei, Z.~Leng, and C.~Li, ``Does role-playing chatbots capture the character personalities? assessing personality traits for role-playing chatbots,'' \emph{arXiv preprint arXiv:2310.17976}, 2023.

\bibitem{lin2023swiftsage}
B.~Y. Lin and et. al., ``Swiftsage: {A} generative agent with fast and slow thinking for complex interactive tasks,'' \emph{arXiv preprint arXiv:2305.17390}, 2023.

\bibitem{jacobs1991}
R.~A. Jacobs and et. al., ``Adaptive mixtures of local experts,'' \emph{Neural computation}, vol.~3, no.~1, pp. 79--87, 1991.

\bibitem{eigen2013learning}
D.~Eigen and et. al., ``Learning factored representations in a deep mixture of experts,'' \emph{arXiv preprint arXiv:1312.4314}, 2013.

\bibitem{du2022glam}
N.~Du and et. al., ``Glam: Efficient scaling of language models with mixture-of-experts,'' in \emph{Proc. ICML}, 2022, pp. 5547--5569.

\bibitem{jaiswal2023leveraging}
R.~K. Jaiswal, M.~Elnourani, S.~Deshmukh, and et. al., ``Leveraging transfer learning for radio map estimation via mixture of experts,'' \emph{Authorea Preprints}, 2023.

\bibitem{lopez2020channel}
L.~M. Lopez-Ramos and et. al., ``Channel gain cartography via mixture of experts,'' in \emph{Proc. IEEE Globecom}, 2020, pp. 1--7.

\bibitem{deng2019next}
B.~Deng, C.~Jiang, H.~Yao, S.~Guo, and S.~Zhao, ``The next generation heterogeneous satellite communication networks: {I}ntegration of resource management and deep reinforcement learning,'' \emph{IEEE Wireless Comm.}, vol.~27, no.~2, pp. 105--111, 2019.

\bibitem{clerckx2023primer}
B.~Clerckx, Y.~Mao, E.~A. Jorswieck, and et. al., ``A primer on rate-splitting multiple access: {T}utorial, myths, and frequently asked questions,'' \emph{IEEE J. Sel. Area Comm.}, 2023.

\bibitem{9575181}
R.~Zhang, K.~Xiong, Y.~Lu, B.~Gao, P.~Fan, and K.~B. Letaief, ``Joint coordinated beamforming and power splitting ratio optimization in {MU}-{MISO} {SWIPT}-enabled {H}et{N}ets: {A} multi-agent ddqn-based approach,'' \emph{IEEE J. Sel. Area. Comm.}, vol.~40, no.~2, pp. 677--693, 2022.

\bibitem{nasir2019multi}
Y.~S. Nasir and D.~Guo, ``Multi-agent deep reinforcement learning for dynamic power allocation in wireless networks,'' \emph{IEEE J. Sel. Area Comm.}, vol.~37, no.~10, pp. 2239--2250, 2019.

\bibitem{ardah2019hybrid}
K.~Ardah and et. al., ``Hybrid analog-digital beamforming design for {SE} and {EE} maximization in massive {MIMO} networks,'' \emph{IEEE Trans. Veh. Technol.}, vol.~69, no.~1, pp. 377--389, 2019.

\bibitem{SemanticRouter}
\BIBentryALTinterwordspacing
Semantic router. 2024. [Online]. Available: \url{https://python.langchain.com/docs/expression_language/cookbook/embedding_router}
\BIBentrySTDinterwordspacing

\bibitem{Enbedding}
\BIBentryALTinterwordspacing
Open ai textual embedding model. 2024. [Online]. Available: \url{https://openai.com/blog/new-and-improved-embedding-model}
\BIBentrySTDinterwordspacing

\bibitem{kasneci2023chatgpt}
E.~Kasneci, K.~Se{\ss}ler, and et. al., ``Chatgpt for good? {O}n opportunities and challenges of large language models for education,'' \emph{Learning and individual differences}, vol. 103, p. 102274, 2023.

\bibitem{shen2018fractional}
K.~Shen and W.~Yu, ``Fractional programming for communication systems-{P}art {I}: {P}ower control and beamforming,'' \emph{IEEE Trans. Signal Process.}, vol.~66, no.~10, pp. 2616--2630, 2018.

\bibitem{ye2023taskexpert}
H.~Ye and D.~Xu, ``Taskexpert: Dynamically assembling multi-task representations with memorial mixture-of-experts,'' in \emph{Proc. IEEE/CVF ICCV}, 2023, pp. 21\,828--21\,837.

\bibitem{zhang2023energy}
R.~Zhang, K.~Xiong, Y.~Lu, P.~Fan, D.~W.~K. Ng, and K.~B. Letaief, ``{E}nergy efficiency maximization in {RIS}-assisted {SWIPT} networks with rsma: {A} ppo-based approach,'' \emph{IEEE J. Sel. Area. Comm.}, vol.~41, no.~5, pp. 1413--1430, 2023.

\bibitem{ren2021probabilistic}
J.~Ren, Y.~Li, Z.~Ding, W.~Pan, and H.~Dong, ``Probabilistic mixture-of-experts for efficient deep reinforcement learning,'' \emph{arXiv preprint arXiv:2104.09122}, 2021.

\bibitem{obando2024mixtures}
J.~Obando-Ceron, G.~Sokar, T.~Willi, C.~Lyle, J.~Farebrother, J.~Foerster, G.~K. Dziugaite, D.~Precup, and P.~S. Castro, ``Mixtures of experts unlock parameter scaling for deep rl,'' \emph{arXiv preprint arXiv:2402.08609}, 2024.

\bibitem{clerckx2019rate}
B.~Clerckx and et. al., ``Rate-splitting unifying {SDMA}, {OMA}, {NOMA}, and multicasting in {MISO} broadcast channel: {A} simple two-user rate analysis,'' \emph{IEEE Wireless Comm. Lett.}, vol.~9, no.~3, pp. 349--353, 2019.

\bibitem{9110869}
C.~Huang, R.~Mo, and C.~Yuen, ``Reconfigurable intelligent surface assisted multiuser {MISO} systems exploiting deep reinforcement learning,'' \emph{IEEE J. Sel. Area Comm.}, vol.~38, no.~8, pp. 1839--1850, 2020.

\bibitem{GPT-3.5}
\BIBentryALTinterwordspacing
The gpt-3.5 model. 2024. [Online]. Available: \url{https://openai.com/blog/gpt-3-5-turbo-fine-tuning-and-api-updates}
\BIBentrySTDinterwordspacing

\bibitem{Langchain}
\BIBentryALTinterwordspacing
The langchain library. 2024. [Online]. Available: \url{https://www.langchain.com/}
\BIBentrySTDinterwordspacing

\bibitem{chen2024benchmarking}
J.~Chen, H.~Lin, X.~Han, and L.~Sun, ``Benchmarking large language models in retrieval-augmented generation,'' in \emph{Proc. AAAI}, vol.~38, no.~16, 2024, pp. 17\,754--17\,762.

\end{thebibliography}
}
\end{CJK}
\end{document}